\documentclass[leqno,12pt]{article}
\usepackage{cite}
\usepackage[a4paper,hscale=0.82,vscale=0.75]{geometry}
\usepackage[parfill]{parskip}
\usepackage{graphicx}
\usepackage{amssymb}

\usepackage{amsmath}
\usepackage{amsthm}
\usepackage{amscd}
\usepackage[all,cmtip]{xy} 
\numberwithin{equation}{section}
\usepackage{empheq}
\usepackage[british]{babel}
\usepackage[small,nohug]{diagrams} 

\title{A framework for structure-preserving submersions\\
and some theorems in relativistic flows
}
\author{Ziyang Hu\footnote{\texttt{z.hu@damtp.cam.ac.uk}}\\
D.A.M.T.P.\\
 University of Cambridge}

\newcommand{\pd}{\partial}
\newcommand{\rs}{\mathbb{R}}

\newcommand{\tr}{\mathrm{Tr\,}}

\DeclareMathOperator{\ad}{Ad}

\DeclareMathOperator{\iprod}{\lrcorner}

\newtheoremstyle{shape0}
  {9pt}
  {9pt}
  {}
  {}
  {\bfseries}
  {.}
  {.5em}
  {}

\newtheoremstyle{shape1}
  {9pt}
  {9pt}
  {\it}
  {}
  {\bfseries}
  {.}
  {.5em}
  {}

\newtheoremstyle{shape2}
  {9pt}
  {9pt}
  {}
  {}
  {\itshape}
  {.}
  {.5em}
  {}

\theoremstyle{shape1}
\newtheorem*{thm*}{Theorem}
\newtheorem*{prop*}{Proposition}

\theoremstyle{shape0}

\theoremstyle{shape2}
\theoremstyle{definition}
\newtheorem*{dfn}{Definition}

\DeclareMathOperator{\rank}{rank}
\DeclareMathOperator{\pr}{pr}

\begin{document}
\maketitle
\abstract{In this paper we first propose a framework of structure-preserving submersions, which generalises the concept of a Riemannian submersion, and dualises the concept of subgeometry, or ``structure-preserving immersions''. The emphasis of our approach is on making precise the free variables and the degree of freedom in a given system, thus making the messy calculations in such problems more bearable and, more importantly, algorithmic. In particular, we derive the degrees of freedom of Riemannian submersions and of Weyl submersions. Then we apply our framework to the study of relativistic dissipationless flow and shear-free flows, obtaining generalisations of the classical Herglotz--Noether theorem to conformally flat spacetime in all dimensions and a partial result of Ellis conjecture to all dimensions.}

\tableofcontents
\newpage

\section*{Introduction}
\addcontentsline{toc}{section}{Introduction}

\subsection{Structure-preserving submersion}

What is a ``structure-preserving submersion''? To answer this question, we first need to know what is meant by a submersion.  Let $M$ and $N$ be smooth manifolds and $f:M\rightarrow N$  a smooth map with constant rank. If $\rank f$, which can be defined as the rank of the Jacobian matrix with respect to {any} local coordinate systems on $M$ and $N$, is equal to $\dim N$, then the map $f$ is said to be a submersion. Of course, when $\rank f=\dim M$, the map is an immersion. It is in this sense that immersions and submersions can be considered the dual of each other.
Even though the definitions involve two manifolds $M$ and $N$ and a map between them, it is usually more intuitive to consider submersions and immersions as occurring on a single manifold. Thus in the case of immersions, the manifold $M$ is thought as a local \emph{submanifold} in $N$, and similarly a submersion is locally considered to be a \emph{foliation} on $M$.

In the case of immersions, geometrical and  physical problems motivates that we consider the geometrical structures on $M$ \emph{induced} from a certain geometrical structure on $N$. For example, if $N$ is equipped with an Riemannian metric, then $M$ is equipped with the induced metric, which is simply the pullback under the map $f$. It is easy to show that this induction of metric structure always exists, and is uniquely defined. However, this is not always the case: for example, the induced structure is not uniquely defined if the structure under consideration is a projective connection. In any case, since it is easy to think in terms of submanifolds, how the induced structure can be defined in a given situation is usually obvious.
A particular point of view is also useful for many problems: sometimes our interest is only focused on what happens on the immersed manifold. In this case, the details of the structure on $N$ is only relevant in an infinitesimal neighbourhood of $f(M)$. Thus, for example, we can define an ``ambient geometry'' of an immersed Riemannian manifold, which is completely determined by the metric on it together with the second fundamental form.

As submersion is the ``dual'' of immersion, what we aim to do in this paper is, by analogy with the above consideration about induction of structures due to immersion, constructing a framework in which we can study the induction of structure due to submersions. Of course, in specific cases, such works have already been done, the most prominent being the the theory of \emph{Riemannian submersions} \cite{BONeill:1966p7043,rsart,ymkkep}: these are Riemannian manifolds $M$ on which we have a number of vector fields defined, and the Lie derivative of the metric under any of these vector fields vanishes. Such vector fields then defines a foliation on the manifold, and the quotient manifold by this foliation is equipped with an Riemannian metric. However, even when restricted to Riemannian submersions, the conventional approach has several shortcomings: in the conventional approach, such a structure is defined, and shown to be completely defined, by a metric on the reduced space $N$ and several ``gluing tensors'' on $M$, together with a set of ``structure equations'' for the submersion \cite{BONeill:1966p7043}. This requires a special coordinate system on $M$, compatible with the foliation, but otherwise the \emph{local symmetries} of the problem is left unclear. Furthermore, an over-emphasis on the use of connections on such spaces makes calculations difficult in such problems, even for relatively simple problems.

The present approach, based on the method of moving frames \cite{WSharpe:1997p5521,AndrewIvey:2003p7876}, is in a sense motivated by the study of ``ambient geometries'' in immersions by considering on the relevant data. In particular, our ``data'' takes into account the local symmetries of the problem: indeed, our approach is fundamentally based on the considerations of these symmetries. As a result, our approach makes clear the functional dependence of the variables in such a theory. At the same time, since the method of moving frames applies in general to any ``generalised geometry'' based on a particular model of homogeneous space, our theory applies in all cases where the submersions preserves such a geometrical structure and is not restricted to metric theories.

Even when studying the problems of Riemannian submersions, our framework presents considerable advantage over the traditional approach based on the theory of connections. Problems related to submersions are notoriously difficult since the number of variables involved is huge, and the variables are almost never independent. Furthermore, one really has to be a veteran in differential geometry in order to be able to successfully obtain useful results by this approach, since when faced with a problem, there is no general strategy for proceeding. By contrast, our method, based on the theory of moving frames, is \emph{algorithmic}, so we always have a strategy for tackling any given problem. Furthermore, the \emph{degree of freedom} of the general theory and particular realisations can be easily calculated using a method based on moving frames \cite{hu1}, so we are no longer dealing with a mess of interrelated quantities whose interrelations we know little about. Also, since we work mainly in the principal bundles, problems involving differential operators in the usual approach become algebraic problems in our approach, and these are much easier to manipulate. Lastly but most importantly, as we shall show, in the principal bundle a structure-preserving submersion essentially amounts to a reduction of the structural group of this bundle, and hence translates to the reduction in the number of independent variables. This will be the key of solving many problems which are traditionally too messy to deal with.

\subsection{Relativistic flows}

We will, in the second part of this paper, discuss applications of our framework of structure-preserving submersions to the study of relativistic flows, especially the flows of Born-rigid (dissipationless) type and of shear-free type. As we will show, that the theory of Born-rigid flow \cite{Giulini:2006p66,pirani1962,born1909} corresponds exactly to Riemannian submersions of  codimension $1$. By using the methods developed in the first part of the paper, we easily obtain results concerning the degrees of freedom of Born-rigid flows of various type. Then we deal with the classical theorem of Herglotz--Noether, which states that in Minkowski spacetime of four dimensions any rotational Born-rigid flow must be an isometry. This theorem, which is classically very messy to prove \cite{Giulini:2006p66,pirani1962,Estabrook:1964p2608,Wahlquist:1966p2548,Wahlquist:1967p2556,herglotz1910,noether1910} and has applications in unexpected places such as in AdS-CFT correspondences \cite{Bhattacharyya:2007p31,me:000}, becomes almost trivial using our approach. Further, it is immediately clear that the theorem also holds for all dimensions and all homogeneous spaces. With a little calculation, we show that for conformally flat spaces, the theorem also holds for all dimensions greater than three, generalising the four dimensional extension by Estabrooks and Wahlquist \cite{Wahlquist:1967p2556}.

Next, we use our method to study flows that still has vanishing shear but possibly non-vanishing expansion, for which the theory of Riemannian submersion is inapplicable, and show that this is modelled exactly by submersions preserving Weyl structures. Then we study a purely geometrical partial result of the celebrated Ellis conjecture \cite{ellis1,LANG,TREC,TREL,ELL1,ELL2,ELL3}, and generalise this result easily to all dimensions.

\subsection{Further applications}

Besides the topic of relativistic fluid treated in this paper, there are other important applications of the theory of structure-preserving submersions. Below we will mention two areas of applications, which will be discussed in further works. 

As we have mentioned, the gist of structure-preserving submersions is reduction of the number of independent variables, and hence obviously the theory has direct application to the problems of dimensional reduction: Kaluza--Klein, de Witt, and Pauli \cite{redgib}. Note that, of these reductions, only the ``old'' Kaluza--Klein theory is treatable using Riemannian submersions, whereas the rest requires the more general theory of structure-preserving submersions. 

As will be discussed in this paper, any isometry of Riemannian spaces is automatically a Riemannian submersion. Thus we can use the framework of structure-preserving submersions to study spaces with symmetries, the obvious benefit being that in this way we have a reduction of independent variables. And by using different models, such studies do not need to be not restricted to isometries. In particular, the theory can be directly applied to the study of black holes of various dimensions, lagrangians with symmetries, and simple quantum systems with symmetrical properties.

\part*{The theory of structure-preserving submersion}
\addcontentsline{toc}{part}{The theory of structure-preserving submersion}

\section{Definition of a structure-preserving submersion}

\subsection{The definition}

We assume the reader is familiar with the language of moving frames, Cartan's generalised geometries and exterior differential systems. See \cite{WSharpe:1997p5521} for review of these topics.

\begin{dfn}
Let $\pr_{M}:P\rightarrow M$ and $\pr_{N}:Q\rightarrow N$ be two Cartan's generalised spaces, namely, $M$ and $N$ are the base manifolds, and $P$ and $Q$ are the principal bundles over $M$ and $N$ respectively. Let $\pi_{i}$, $i=1,\dots,\dim N$ and $\pi_{\mu}$ be the Cartan connection on $Q$, in which the $\pi_{i}$ are the horizontal forms and $\pi_{\mu}$ are the vertical forms. Let $\omega_{i}$, $i=1,\dots,\dim N$, $\omega_{a}$, $a=\dim N+1,\dots,\dim M$ and $\omega_{\alpha}$ be the Cartan connection on $P$, in which the $\omega_{i}$ and $\omega_{a}$ are the horizontal forms and the $\omega_{\alpha}$ are the vertical forms. A structure-preserving submersion is a solution (i.e., an \emph{integral variety}) of the exterior differential system
\begin{equation}
  \label{eq:sbeds}
  \pi_{i}=\omega_{i}
\end{equation}
with the independence conditions given by the forms
\[
\pi_{i},\qquad\pi_{\mu},\qquad\omega_{a}
\]
together with those of the forms
\[
\omega_{\alpha}
\]
corresponding to the Lie algebra having trivial actions on the forms $\pi_{i}$, and having the space of forms $\omega_{a}$ as an invariant subspace.
\end{dfn}

This definition is formulated such that it is as concise as possible and if we accept it, we can rapidly do calculations on a structure-preserving submersion without discussing many subtle points about structure-preserving submersions. As a price to pay, the definition is not very intuitive. 

Let us first check that it is, first and foremost, a submersion. First, generically, assume that $f:A\rightarrow B$ is a submersion, then we can form the graph of this map, which is a submanifold of $S\subset A\times B$. It is clear that $\dim S=\dim A$. If we have a coframe $\omega_{A}$ on $A$ and $\omega_{I}$ on $B$, then $\omega_{A}$, $\omega_{I}$ together form a coframe on $A\rightarrow B$. As the submanifold $S$ arises from the function $f$, the forms $\iota^{*}\omega_{A}$ are independent one-forms on $S$ where $\iota:S\rightarrow A\times B$ is the canonical inclusion map. On the other hand, the rank condition of the submersion means that $\iota^{*}\omega_{A}$ are also independent one-forms on $S$. Conversely, if these conditions on the forms are satisfied, then the submanifold $S$ arises locally as the graph of a submersion map.

For our problem, it is thus clear that the solution of the differential system \eqref{eq:sbeds} arises from a submersion for which the map is $f:A\rightarrow B$, and $B=Q$. The manifold $A$ is a little bit more complicated: it has first of all the coframe $\pi_{i},\pi_{\mu},\omega_{a}$ and $\omega^{*}_{\alpha}$, where the star over $\omega_{\alpha}$ meaning that only those corresponding to trivial actions on $\pi_{i}$ and preserving the subspace of $\omega_{a}$ are included. It is also a sub-bundle of the principal bundle $P$: notice that the forms $\pi_{i}$ and $\omega_{a}$ can be taken as a set of horizontal forms.

If, instead of dealing with the principal bundles $P$ and $Q$, we deal with sections on them, and assume that for a certain section the definition still holds when we substitute the forms with the pullbacks of forms onto the section, with the independent conditions now only given by $\pi_{i}$ and $\omega_{a}$, since the vertical forms for any section are expressible linearly in terms of the pullbacks of the horizontal forms. Then we see that under this section, the integral variety we have found corresponds to a submersion $M\rightarrow N$.

To summarise, the integral variety we have found corresponds to the following situation:
\[  \xymatrix{
M\times H\ar[r]^{\approx}\ar[rd]_{\pr_{1}}&A\ar[r]^{f}\ar[d]_{\pr_{M}}&Q\ar[d]^{\pr_{N}}&N\times H_{N}\ar[l]_{\approx}\ar[ld]^{\pr_{1}}\\
&M\ar[r]^{\sigma}&N&
}
\]
in the diagram above, $\sigma:M\rightarrow N$ is a submersion on the base manifolds $M$ and $N$, and we have found a submersion $f$ covering $\sigma$ from a certain principal bundle $A$ over $M$ to the principal $Q$ over $N$.

What is the principal bundle $A$ and what is the group $H$? Since the map $f$ is a submersion, it is clear that the group $H_{N}$ is a subgroup of the group $H$. This is also clearly seen from the fact that $\omega_{\mu}$ is the Maurer-Cartan forms when restricted to a vertical subspace, which is isomorphic to $H_{N}$, and $\omega_{\mu}$, $\omega^{*}_{\alpha}$ together can be taken as the Maurer-Cartan forms on $H$. The forms $\omega^{*}_{\alpha}$ are practically found as follows: for any Cartan connection $\omega_{A}$ of a principal bundle with Lie algebra $\mathfrak g$, the transformation under the right action of the principal group itself is
\[
h:\omega_{A}\rightarrow \ad_{h}(\omega_{A}),\qquad h\in \mathfrak g.
\]
This is the equation that we can use to determine what are the forms $\omega^{*}_{\alpha}$. For example, if the principal bundle $P$ and $Q$ both correspond to the principal bundle for geometries with a projective connection, then we can write the Cartan connection on $P$ as a matrix
\[
\begin{pmatrix}
  \omega_{00}&\omega_{0i}&\omega_{0a}\\
  \omega_{i}&\omega_{ij}&\omega_{ia}\\
  \omega_{a}&\omega_{ai}&\omega_{ab}
\end{pmatrix}
\]
where not all forms are independent: in particular, since we are dealing with projective geometry, $\omega_{00}+\omega_{11}+\dots =0$.
 Under the principal right action, the above connection matrix is transformed to
\[
\begin{pmatrix}
  h_{00}&h_{0i}&h_{0a}\\
  0&h_{ij}&h_{ia}\\
  0&h_{ai}&h_{ab}
\end{pmatrix}
\begin{pmatrix}
  \omega_{00}&\omega_{0i}&\omega_{0a}\\
  \omega_{i}&\omega_{ij}&\omega_{ia}\\
  \omega_{a}&\omega_{ai}&\omega_{ab}
\end{pmatrix}
-
\begin{pmatrix}
  \omega_{00}&\omega_{0i}&\omega_{0a}\\
  \omega_{i}&\omega_{ij}&\omega_{ia}\\
  \omega_{a}&\omega_{ai}&\omega_{ab}
\end{pmatrix}
\begin{pmatrix}
  h_{00}&h_{0i}&h_{0a}\\
  0&h_{ij}&h_{ia}\\
  0&h_{ai}&h_{ab}
\end{pmatrix}.
\]
This means that we have one equation of the form
\begin{equation}
  \label{eq:freeforms}
\begin{pmatrix}
  h_{00}&h_{0i}&h_{0a}\\
  0&h_{ij}&h_{ia}\\
  0&h_{ai}&h_{ab}
\end{pmatrix}
\begin{pmatrix}
  0&0&0\\
  \omega_{i}&0&0\\
  0&0&0
\end{pmatrix}
-
\begin{pmatrix}
  0&0&0\\
  \omega_{i}&0&0\\
  0&0&0
\end{pmatrix}
\begin{pmatrix}
  h_{00}&h_{0i}&h_{0a}\\
  0&h_{ij}&h_{ia}\\
  0&h_{ai}&h_{ab}
\end{pmatrix}=
\begin{pmatrix}
  0\\
  0\\
  0
\end{pmatrix}
\end{equation}
and
\begin{equation}
  \label{eq:freeforms2}
\begin{pmatrix}
  h_{00}&h_{0i}&h_{0a}\\
  0&h_{ij}&h_{ia}\\
  0&h_{ai}&h_{ab}
\end{pmatrix}
\begin{pmatrix}
  0&0&0\\
  0&0&0\\
  \omega_{a}&0&0
\end{pmatrix}
-
\begin{pmatrix}
  0&0&0\\
  0&0&0\\
  \omega_{a}&0&0
\end{pmatrix}
\begin{pmatrix}
  h_{00}&h_{0i}&h_{0a}\\
  0&h_{ij}&h_{ia}\\
  0&h_{ai}&h_{ab}
\end{pmatrix}=
\begin{pmatrix}
  0\\
  0\\
  \star
\end{pmatrix}
\end{equation}
for each $\omega_{i}$, $i=1,2,\dots,\dim N$. These will tell us which linear combinations of $h_{00}$, $h_{0i}$, $h_{0a}$, $h_{ij}$, $h_{ia}$, $h_{ai}$ and $h_{ab}$ need to be set to known constants (zero). The complement of those that are set to constants gives the linear combinations of the forms that are retained in the bundle $A$.

Thus, we see that the group $H$ satisfies $H_{N}\subset H\subset H_{M}$, where the subset symbol means subgroup, and is uniquely determined by the procedure above. Another way of saying the same thing is that $A$ is obtained from $P$ by a \emph{reduction of the principal bundle} from the structural group $H_{M}$ to $H$. Intuitively, the significance of this reduction is as follows: for any structure-preserving submersions, the horizontal one-forms $\omega_{i}$ are replaced with the one-forms $\pi_{i}$ arising from the submersion. The $\pi_{i}$ satisfies its own structural equations (and hence its structure is ``preserved'' in $M$), and consequently any right action in $P$ that ``moves'' $\pi_{i}$ in any non-trivial way is forbidden. Or, in the language of moving frames (instead of coframes), a part of the frame is already fixed, so any transformation of the frames not preserving completely this part of the frame is no longer allowed. Yet another way of saying the same thing is: we have the frame on $M$, but also the frame deduced from the one on $N$, which can be interpreted as a partial frame on $M$. We need to use some of the degree of freedom of the group $H_{M}$ in order to \emph{align} the frame on $M$ with the partial frame on $N$, and hence after this alignment, some of the degree of freedom of $H_{M}$ is lost and we obtain the subgroup $H$.
The reason that we also disallow mixing of $\omega_{a}$ and $\pi_{i}$ is that, we know that for any principal bundle, the right action of the group just changes the choice of identity of the group in the bundle and should not have any real effect. However, if $\omega_{a}$ can be changed into $\pi_{i}$ by such an action, then the definition of structure-preserving submersion will depend on such a choice, and hence our definition would make no sense.

Thus, in summary, we have the following commutative diagram:
\[  \xymatrix{
M\times H_{M}\ar[d]_{\simeq}&M\times H\ar[l]_{\tilde\rho}\ar[r]^{\tilde f}\ar[d]_{\simeq}&N\times H_{N}\ar[d]^{\simeq}\\
P\ar[d]_{\pr_{M}}&A\ar[l]_{\rho}\ar[r]^{f}\ar[d]_{\pr_{M}}&Q\ar[d]^{\pr_{N}}\\
M&M\ar[l]_{=}\ar[r]^{\sigma}&N
}
\]
where $\rho$ is the inclusion map arising from the reduction of the principal bundle (note the direction of arrow). From this diagram, we also have the following interpretation of the integral variety in our definition: using the maps $f$ and $\rho$, we pull back the coframes on $P$ and $Q$ to $A$, and the independent forms form a co-frame on $A$. The map $f$, which covers the submersion $\sigma$, is the \emph{structure-preserving} submersion.

Note that our definition does not explicitly state that the group $H_{N}$ must be a subgroup of $H_{M}$, but if this is not the case, it is impossible to find any solution of the required differential system: there must be non-trivial relations among the $\pi_{\mu}$. If this case arises in applications, we need to start again and try to find structure-preserving submersions preserving a subgroup of $H_{N}$.

Note also it is in general impossible to define a covering submersion map directly from $P$ to $Q$: this requires us finding a submersion from the group $H_{M}$ to $H_{N}$, which also \emph{preserves the group structure}, i.e., the map must be a surjective homomorphism. Such maps do not in general exist, even when $H_{N}$ is a subgroup of $H_{M}$. Thus we see that the reduction of the principal bundle $\rho$ is essential.

On the other hand, our definition, which only explicitly talks about the exterior differential system \eqref{eq:sbeds}, makes all these discussions about reductions of bundles, etc., redundant, even though they are certainly helpful for an intuitive understanding. 

\subsection{The structural equations}
The use of coframes, which is more general than connections, leads us naturally to the equivalence problem. From the general theory of equivalence problems defined by coframes (see \cite{Fels:1998p7335} for a review of the equivalence problem using moving frames), we know that the \emph{structural equations} of the coframe contains all the differential invariants of the problem. For any structure-preserving submersions, our definition already gives us a co-frame, namely the set
\begin{equation}
  \label{eq:ind}
  \pi_{i},\qquad\pi_{\mu},\qquad\omega_{a},\qquad\omega_{\alpha}^{*}.
\end{equation}
The problem \emph{may} also give us additional functions that must be included in the determination of the equivalence problem, though such cases are rare.
Let us now study, without specialising to specific groups, their structural equations.

The structural equations of structure-preserving submersions are deduced from four sets of equations: the first set is the structural equations on $N$:
\begin{equation}
  \label{eq:s1}
  d\pi_{i}=\cdots,\qquad d\pi_{\mu}=\cdots,
\end{equation}
the second set is the structural equations on $M$:
\begin{equation}
  \label{eq:s2}
  d\omega_{i}=\cdots,\qquad d\omega_{a}=\cdots,\qquad d\omega_{\alpha}=\cdots,
\end{equation}
the third set is the exterior differential system itself:
\begin{equation}
  \label{eq:s3}
  \omega_{i}=\pi_{i},\qquad d\omega_{i}=d\pi_{i},
\end{equation}
and finally, the fourth set is the decomposition of the $\omega_{\alpha}$ that are not independent, which we write as $\omega_{\alpha}^{\dag}$:
\begin{equation}
  \label{eq:s4}
  \omega_{\alpha}^{\dag}=A_{\alpha i}\pi_{i}+B_{\alpha a}\omega_{a}+C_{\alpha\mu}\pi_{\mu}+D_{\alpha\beta}\omega_{\beta}^{*},\qquad d\omega_{\alpha}^{\dag}=\cdots.
\end{equation}

These equations are not all independent: due to the dimension of the integral variety, we know that from them we should deduce a set of independent structural equations, on the left hand sides of which are the exterior derivatives of the independent one-forms \eqref{eq:ind}. On the other hand, all functions appearing on the right hand sides of \eqref{eq:s1}, \eqref{eq:s2}, \eqref{eq:s3}, \eqref{eq:s4} must be taken as differential invariants of the system. The redundant equations among them then gives the \emph{algebraic relations} among the differential invariants.

The first equations of \eqref{eq:s3} and \eqref{eq:s4} are the only one-form equations in our system. We should immediately use them to substitute all occurrences of $\omega_{i}$ and $\omega_{\alpha}^{\dag}$ with the independent forms. Once this is done, we see that both \eqref{eq:s1} and \eqref{eq:s2} contains expressions for $d\pi_{i}$, both \eqref{eq:s3} and \eqref{eq:s4} contains expressions for $d\omega^{\dag}_{\alpha}$. Using these equalities, we obtain all constraints of the system at this level. These constraints may make all differential invariants that do not occur explicitly in the new coframe structural equations whose left hand sides are the exterior derivatives of \eqref{eq:ind} completely expressible in terms of the invariants that occur explicitly in the new coframe structural equations. If this is not the case, then any invariants that do not explicitly occur at this level must be included as additional scalar functions in the equivalence problem, which shows that the co-frame does not uniquely determine the structure-preserving submersion. This is analogous to the case in immersion where the induced structure on the submanifold is not uniquely defined.

The relations among the differential invariants obtained above will be called the \emph{defining relations} for them. There are two other kinds of relations for them. 

One kind of such relations is called the \emph{generic relations} and they involve the \emph{coframe derivatives} of the differential invariants. For example, if $I$ is any invariant, possibly with indices, then its coframe derivative is defined as
\begin{equation}
  \label{eq:gen}
  dI=I_{;i}\pi_{i}+I_{;a}\omega_{a}+I_{;\mu}\omega_{\mu}+I_{;\alpha}\omega_{\alpha}^{*},
\end{equation}
but this defining equation can be immediately differentiated again, which may generate higher order coframe derivatives. The generic relations are just the relations $d^{2}I=0$. Note that
the coframe derivatives $I_{;i}, I_{;a}, I_{\mu}, I_{\alpha}$ are considered algebraically \emph{independent} quantities unless there is an explicit relation for them, which.

The next kind is called the \emph{Bianchi relations}: these are obtained by exterior differentiating \eqref{eq:s1}, \eqref{eq:s2}, \eqref{eq:s3}, \eqref{eq:s4} and use the identity $d^{2}=0$. Deriving these relations is usually a rather tedious process, so it is important to note the following in order to reduce unnecessary work: for example, we can calculate $d^{2}\pi_{i}$ by either exterior differentiating the equation in \eqref{eq:s2}, or the equation in \eqref{eq:s3}, but \emph{we only need to differentiate one of them} since they imply each other. Indeed, we have already obtained relations for the invariants which makes the relation $d\pi_{i}=d\omega_{i}$ and identity, so $d^{2}\pi_{i}=d^{2}\omega_{i}=0$ is an identity as well. The same reasoning applies to the equations involving $\omega_{\alpha}^{\dag}$.

The last kind of these relations is called the \emph{derived relations}. For example, let $f(I)=0$ be an algebraic relation (zero form equation) of the previous kinds. Then a relation can be obtained by exterior differentiating: $df(I)=0$. Of course, derived relations can be further derived to obtain an infinite tower of relations, but if we truncate the tower of differential invariants by only considering invariants whose number of indices is less than a given number, then the total number of relations at this stage is finite.

If at any stage a relation we obtain is \emph{incompatible}, for example of the form $1=0$, this simply means that no required integral variety exists. In particular, it is easy to show by considering the structural constants of the groups that, if $H_{N}$ is not a subgroup of $H_{M}$, then incompatibility will occur.

\section{Riemannian submersion}

We now give our first example of a consistent structure-preserving submersion. This is none other than the case of Riemannian submersion. It should be clear from our exposition below how similar procedures can be applied to other groups whenever the submersion is consistent, and when something that we do is specific for Riemannian structures we will mention so explicitly.

\subsection{The structure of a Riemannian submersion}

Here the appropriate group is $H_{M}=SO(m)$, $H_{N}=SO(p)$. We assume $m>p$: the case $m=p$ amounts to a rather awkward study of isometries. In matrix notation, the connection on $Q$ is
\begin{equation}
  \label{eq:conq}
  \begin{pmatrix}
  0&0\\
  \pi_{i}&\pi_{ij}
\end{pmatrix},\qquad \pi_{ij}=-\pi_{ji}.
\end{equation}
and the connection on $P$ is
\begin{equation}
  \label{eq:conp}
\begin{pmatrix}
0&0&0\\
\omega_{a}&\omega_{ab}&\omega_{ai}\\
\omega_{i}&\omega_{ia}&\omega_{ij}
  \end{pmatrix},\qquad \omega_{ab}=-\omega_{ba},\quad\omega_{ij}=-\omega_{ji},\quad\omega_{ai}=-\omega_{ia}.
  \end{equation}
The $Q$ structural equations are
\begin{equation}
  \label{eq:strq}
    \left\{
    \begin{aligned}
      d\pi_{i}&=-\pi_{ij}\wedge\pi_{j},\\
      d\pi_{ij}&=-\pi_{ik}\wedge\pi_{kj}+\tfrac{1}{2}S_{ijkl}\,\pi^{k}\wedge\pi^{l}.
    \end{aligned}
  \right.
\end{equation}
And the $P$ structural equations are
\begin{equation}
  \label{eq:strp}
    \left\{
    \begin{aligned}
      d\omega_{\mu}&=-\omega_{\mu\nu}\wedge\omega_{\nu},\\
      d\omega_{\mu\nu}&=-\omega_{\mu\lambda}\wedge\omega_{\lambda\nu}+\tfrac{1}{2}R_{\mu\nu\rho\lambda}\,\omega^{\rho}\wedge\omega^{\lambda},
    \end{aligned}
  \right.
\end{equation}
where the index $\mu$ runs over both the ranges $i$ and $a$.

Let us find the independent forms on the reduced principal bundle: these will obviously include $\pi_{i},\omega_{a}$ and $\pi_{ij}$. On the other hand, the action \eqref{eq:freeforms} is now just the linear action on vector spaces. Thus we require
\[
\begin{pmatrix}
  h_{ab}&h_{ai}\\
  h_{ia}&h_{ij}
\end{pmatrix}
\begin{pmatrix}
  0\\
  \star
\end{pmatrix}
=0,
\]
where $\star$ can be anything. This requires $h_{ai}=0$, $h_{ij}=0$, and by the symmetry of the Lie algebra, $h_{ia}=0$ as well. Note that the Lie subalgebra obtained thus preserves the subspace of the $\omega_{a}$ as well,  hence $\omega_{ab}$ are  independent forms on the reduced bundle.

For the non-independent forms, $\omega_{i}=\pi_{i}$. For $\omega_{ai}=-\omega_{ia}$, we can write
\[
\omega_{ai}=-\omega_{ia}=K_{iab}\omega_{b}-M_{ija}\pi_{j}+A_{aibc}\omega_{bc}+B_{aijk}\pi_{jk}.
\]
Let us use the equation $d\pi_{i}=d\omega_{i}$ immediately. This gives $\pi_{ij}\wedge\pi_{j}=\omega_{ij}\wedge\pi_{j}+\omega_{ia}\wedge\omega_{a}$, which is
\[
\pi_{ij}\wedge\pi_{j}=\omega_{ij}\wedge\pi_{j}-(K_{iab}\omega_{b}-M_{ika}\pi_{k}+A_{aibc}\omega_{bc}+B_{aijk}\pi_{jk})\wedge\omega_{a}.
\]
On the right hand side, all terms not involving $\pi_{i}$ need to vanish. This gives
\[
A_{aibc}=0,\qquad B_{aijk}=0,\qquad K_{iab}=K_{iba}.
\]
With this, we have
\[
(\pi_{ij}-\omega_{ij}+M_{ija}\omega_{a})\wedge\pi_{j}=0,
\]
which shows that we need to have
\[
\omega_{ij}=\pi_{ij}+M_{ija}\omega_{a}+C_{ijk}\pi_{k},\qquad C_{ijk}=C_{ikj}.
\]
However, the indices on $\omega_{ij}$ and $\pi_{ij}$ are both antisymmetric. Symmetrising, we have
\[
M_{(ij)a}\omega_{a}+C_{(ij)k}\omega_{k}=0,
\]
hence
\[
M_{ija}=-M_{jia},\qquad C_{ijk}=-C_{jik},
\]
but $C_{ijk}$ is antisymmetric in the first two indices while symmetric in the last two indices, hence it is zero, $C_{ijk}=0$. \emph{This property is due to the property of the orthogonal group}: if we use the general linear group, there will be a differential invariant $C_{ijk}$. 

Now we know how to express the dependent forms in terms of the independent ones:
\[
\left\{
  \begin{aligned}
    \omega_{i}&=\pi_{i},\\
    \omega_{ij}&=\pi_{ij}+M_{ija}\omega_{a},\\
    \omega_{ai}&=K_{iab}\omega_{b}-M_{ija}\pi_{j},
  \end{aligned}
\right.\qquad M_{ija}=-M_{jia},\qquad K_{iab}=K_{iba}.
\]

\subsection{The connection}

Riemannian geometry is a ``nice'' geometry in the sense that all of its differential invariants are \emph{tensors}. This means that in the coframe derivatives given by \eqref{eq:gen}, only those coframe derivatives with respect to the horizontal forms are algebraically independent. We can prove this easily: expanding the Bianchi relation for $d^{2}\pi_{i}=0$, we have
\begin{align*}
  0{}={}&\tfrac{1}{2}S_{ijkl;m}\pi_{m}\wedge\pi_{k}\wedge\pi_{l}+\tfrac{1}{2}S_{ijkl;\underline{mn}}\pi_{mn}\wedge\pi_{k}\wedge\pi_{l}\\
  &-\tfrac{1}{2}S_{iklm}\pi_{l}\wedge\pi_{m}\wedge\pi_{kj}+\tfrac{1}{2}S_{kjlm}\pi_{ik}\wedge\pi_{l}\wedge\pi_{m}\\
  &-\tfrac{1}{2}S_{ijkl}\pi_{km}\wedge\pi_{m}\wedge\pi_{l}+\tfrac{1}{2}S_{ijkl}\pi_{k}\wedge\pi_{lm}\wedge\pi_{m},
\end{align*}
the underlined indices $\underline{mn}$ meaning that it should be considered a \emph{single} index: it is the derivation index with respect to $\pi_{mn}$. If we focus on the terms involving forms like $\pi_{mn}\wedge\pi_{k}\wedge\pi_{l}$, we see that $S_{ijkl;\underline{mn}}$ is expressed linearly in terms of $S_{ijkl}$ itself, which is what we want to show. But this expression is just the transformation of a tensor quantity with the indices $ijkl$ in the principal bundle $Q$! By the same argument, $S_{\mu\nu\rho\lambda}$ transforms as a tensor in $P$. These immediately imply that they both transform as tensor quantities in the bundle $A$ by considering the effects of the pullbacks.

Using entirely similar argument, by considering the relation $d^{2}\omega_{ia}=0$ for example, it is straightforward, though a bit tedious, to show directly that the functions $M_{ija}$ and $K_{iab}$ transforms as tensors in the bundle $A$, with all their indices tensor indices. This justifies, \emph{a posteriori}, our calling the forms \eqref{eq:ind} the \emph{connection} on the bundle $A$. If we denote the connection on $A$ by $\nabla$, we see that for a tensor $a_{ai}$, we have
\[
\nabla a_{ai}\equiv a_{ai;j}\,\omega_{j}+a_{ai;b}\,\omega_{b}=da_{ai}+\omega_{ab}\,a_{bi}+\pi_{ij}\,a_{aj}.
\]
The transformation laws for all other indices can be deduced by extending this formula linearly. {This connection \emph{is not the same} as the Levi-Civita connection on the bundle $P$}: indeed, it is related to the Levi-Civita connection by the relation we already derived
\[
\omega_{ij}=\pi_{ij}+M_{ija}\omega_{a}.
\]
As we will see shortly, it is under the connection on $A$ that any quantities that are invariant on each leaf of the foliation (for example $S_{ijkl}$) have all their covariant derivatives with respect to the $a$ indices vanishing. If we use the connection deduced directly from the Levi-Civita connection on $P$, we have a much more difficult situation.

We can augment this connection with the following construction. From the structural equations on $A$, we have $d\pi_{i}=-\pi_{ij}\wedge\pi_{j}$, so the distribution defined by $\pi_{i}=0$ is completely integrable: the integral varieties are just the fibres of the submersion, and every fibre is projected into a single point of $M$ under the submersion.
\[
      d\omega_{ab}=-\omega_{ac}\wedge\omega_{cb}-\omega_{ai}\wedge\omega_{ib}+\tfrac{1}{2}R_{abcd}\,\omega_{c}\wedge\omega_{d}+R_{abci}\,\omega_{c}\wedge\omega_{i}+\tfrac{1}{2}R_{abij}\,\omega_{i}\wedge\omega_{j},
\]
which, after expansion of the non-independent forms,
\[
      d\omega_{ab}=-\omega_{ac}\wedge\omega_{cb}+K_{iac}K_{ibd}\,\omega_{c}\wedge\omega_{d}+\tfrac{1}{2}R_{abcd}\,\omega_{c}\wedge\omega_{d}\pmod{\pi_{i}}.
\]
Hence if we define
\begin{equation}
  \label{eq:fibrecurv}
  S_{abcd}=R_{abcd}+K_{iac}K_{ibd}-K_{iad}K_{ibc}
\end{equation}
then $S_{abcd}$ is the Riemannian curvature tensor on the fibres. It is algebraically equivalent to $R_{abcd}$, so in principle we can take either set to be independent, but practically it is obviously much better to take $S_{abcd}$ due to its interpretation. Unlike $S_{ijkl}$, it can vary in both the $\omega_{i}$ and $\omega_{a}$ directions.

Thus, schematically, the connection $\nabla$ on $A$ splits into two parts, one part $\pi_{ij}$ gives the curvature on $N$, whereas the other part $\omega_{ab}$ gives the curvature on the fibres. It is, however, the ``gluing data'' $M_{ija}$, $K_{iab}$ that is most interesting for the structure-preserving submersion.

\subsection{The curvature}
The second equation of \eqref{eq:strq} gives the Riemannian curvature of the space $N$, and all the forms appearing in this equation are independent. The second equation of \eqref{eq:strp} gives the Riemannian curvature of the space $M$, but many of the forms appearing on either side are not independent. Substituting with the independent forms, we first have, for the {Equations for} $\omega_{ia}$:
\[
d\omega_{ia}{}={}-\omega_{ij}\wedge\omega_{ja}-\omega_{ib}\wedge\omega_{ba}+\tfrac{1}{2}R_{iajk}\,\omega_{j}\wedge\omega_{k}+R_{iajb}\,\omega_{j}\wedge\omega_{b}+\tfrac{1}{2}R_{iabc}\,\omega_{b}\wedge\omega_{c}.
\]
The left hand side gives
\begin{align*}
  d\omega_{ia}{}={}&d(M_{ija}\,\omega_{j}-K_{iab}\,\omega_{b})\\
  {}={}&M_{ija;b}\,\omega^{b}\wedge\omega^{j}+M_{ija;k}\,\omega^{k}\wedge\omega_{j}
  -K_{iab;c}\,\omega_{c}\wedge\omega_{b}-K_{iab;j}\,\omega_{j}\wedge\omega_{b}\\
  &-M_{kja}\,\pi_{ik}\wedge\omega_{j}-M_{ika}\,\pi_{jk}\wedge\omega_{j}-M_{ijc}\,\omega_{ac}\wedge\omega_{j}\\
  &+K_{kab}\,\pi_{ik}\wedge\omega_{b}+K_{icb}\,\omega_{ac}\wedge\omega_{b}+K_{iac}\,\omega_{bc}\wedge\omega_{b}\\
  &+M_{ija}(-\pi_{jk}\wedge\omega_{k})-K_{iab}(-\omega_{bc}\wedge\omega_{c}-K_{jbc}\,\omega_{c}\wedge\omega_{j}-M_{ijb}\,\omega_{i}\wedge\omega_{j}).
\end{align*}
while the right hand side gives
\begin{align*}
  d\omega_{ia}{}={}&K_{ibc}\,\omega_{c}\wedge\omega_{ba}-M_{ikb}\,\omega_{k}\wedge\omega_{ba}\\
  &+K_{jac}\pi_{ij}\wedge\omega_{c}-M_{jka}\,\pi_{ij}\wedge\omega_{k}+M_{ijb}K_{jac}\,\omega_{b}\wedge\omega_{c}-M_{ijb}M_{jka}\,\omega_{b}\wedge\omega_{k}\\
  &+\tfrac{1}{2}R_{iabc}\,\omega_{b}\wedge\omega_{c}+R_{iajb}\,\omega_{j}\wedge\omega_{b}+\tfrac{1}{2}R_{iajk}\,\omega_{j}\wedge\omega_{k}.
\end{align*}
Equating the two sides, all terms containing $\omega_{ab}$ or $\pi_{ij}$ cancel (this is because we already know that all zero-forms in the expression are tensors). The rest gives three relations
\begin{align*}
  R_{aibc}&=-K_{iab;c}+K_{iac;b}-M_{kib}K_{ack}+M_{kic}K_{kab},\\
  R_{aibj}&=M_{ikb}M_{jka}-M_{ija;b}-K_{iab;j}-K_{iac}K_{jbc},\\
  R_{aijk}&=M_{ija;k}-M_{ika;j}-2M_{jkb}K_{iab}.
\end{align*}
We can also calculate $d\omega_{ij}$ to obtain further sets of such relations. However, since $\omega_{ab}$ is \emph{independent}, we do not get such relations for this equation, though we have already defined $S_{abcd}$ by \eqref{eq:fibrecurv}.
Thus, we have the following ``dictionary'' collecting what we obtain from above:
\begin{equation}
  \label{eq:oldriemcurv}
  \left\{
    \begin{aligned}
      R_{abcd}&=S_{abcd}-K_{iac}K_{ibd}+K_{iad}K_{ibc},\\
      R_{ijkl}&=S_{ijkl}+M_{ila}M_{jka}-M_{ika}M_{jla}-2M_{ija}M_{kla},\\
      R_{ijab}&=M_{ikb}M_{jka}-M_{ika}M_{jkb}-M_{ija;b}+M_{ijb;a}+K_{jac}K_{ibc}-K_{iac}K_{jbc},\\
      R_{ijkb}&=M_{ijb;k}-M_{jka}K_{iab}+M_{ika}K_{jab}+M_{ija}K_{kab},\\
      R_{aibc}&=-K_{iab;c}+K_{iac;b}-M_{kib}K_{ack}+M_{kic}K_{kab},\\
      R_{aibj}&=M_{ikb}M_{jka}-M_{ija;b}-K_{iab;j}-K_{iac}K_{jbc},\\
      R_{aijk}&=M_{ija;k}-M_{ika;j}-2M_{jkb}K_{iab}.
    \end{aligned}
  \right.
\end{equation}
It looks as if the left hand sides contain all components of the Riemannian tensor for the space $M$, but actually at this stage we can only be sure of the symmetries $R_{\mu\nu\rho\lambda}=-R_{\nu\mu\rho\lambda}=-R_{\mu\nu\lambda\rho}$. We need the Bianchi identity $R_{\mu[\nu\rho\lambda]}=0$ to really obtain all components of the Riemannian tensor $R_{\mu\nu\rho\lambda}$. We will discuss Bianchi relations later, but by our previous discussion about the general case, we know that the Bianchi identities for $R_{\mu\nu\rho\lambda}$ can be deduced from the Bianchi relations for the coframe on $A$, and hence we can now be certain that the differential invariants $S_{ijkl}$, $S_{abcd}$, $M_{ija}$, $K_{iab}$ alone completely determine the geometry of the submersion: there are no additional functions to be considered in addition to the coframe.

Thus, eliminating the quantities $R_{\mu\nu\rho\lambda}$, our structural equations for the coframe are now written as
  \begin{equation}
  \label{eq:redrsreq}
  \left\{
    \begin{aligned}
      d\pi_{i}{}={}&-\pi_{ij}\wedge\pi_{j},\\
      d\omega_{a}{}={}&-\omega_{ab}\wedge\omega_{b}-K_{iab}\,\omega_{b}\wedge\pi_{i}-M_{ija}\,\pi_{i}\wedge\pi_{j},\\
      d\pi_{ij}{}={}&-\pi_{ik}\wedge\pi_{kj}+\tfrac{1}{2}S_{ijkl}\,\pi_{k}\wedge\pi_{l},\\
      d\omega_{ab}{}={}&-\omega_{ac}\wedge\omega_{cb}+\tfrac{1}{2}S_{abcd}\,\omega_{c}\wedge\omega_{d}\\
      &-2K_{ic[a;b]}\,\omega_{c}\wedge\pi_{i}+\tfrac{1}{2}(-2M_{ij[a;b]}-K_{iac}K_{jbc}+K_{ibc}K_{jac})\pi_{i}\wedge\pi_{j}.
    \end{aligned}
  \right.
\end{equation}
with the defining relations
\begin{equation}
  \label{eq:defrel}
  \begin{gathered}
    R_{ijkl}=-R_{jikl}=-R_{ijlk},\qquad R_{abcd}=-R_{abdc}=-R_{bacd},\\
    M_{ija}=-M_{jia},\qquad K_{iab}=+K_{iba}.
  \end{gathered}
\end{equation}
These are the structural relations for Riemannian submersions. From now on we will study the submersions by considering only consequences of \eqref{eq:redrsreq}, and do not talk about the bundle $P$ or $Q$ or the invariants $R_{\mu\nu\rho\lambda}$ any more.

\subsection{Algebraic relations for the invariants}
\label{sec:algebraic-relations}

What are conventionally called the ``first Bianchi identities'' are obtained by the relations $d^{2}\pi_{i}=0$ and $d^{2}\omega_{a}=0$. The first gives the usual $S_{i[jkl]}=0$, whereas the second gives
\begin{align*}
  S_{a[bcd]}&=0,\\
  M_{ija;k}+M_{jka;i}+M_{kia;j}&=M_{ijb}K_{kab}+M_{jkb}K_{iab}+M_{kib}K_{jab},\\
  -K_{iab;j}+K_{jab;i}&=M_{ija;b}+M_{ijb;a},
\end{align*}

The ``second Bianchi identities'' are obtained from the relations $d^{2}\pi_{ij}=0$ and $d^{2}\omega_{ab}=0$, which gives
\begin{align*}
  S_{ij[kl;m]}&=0,\\
  S_{ijkl;a}&=0,\\
  S_{ab[cd;e]}&=0,\\
  S_{abcd;i}&=2A_{abci;d}-S_{abe[d}K_{|i|c]e}-S_{abde}K_{iec},\\
  A_{ab[ij;k]}&=2A_{abc[i}M_{jk]c},\\
  A_{abc[i;j]}&=-\tfrac{1}{2}A_{abij;c}-A_{abd[i}K_{j]dc}-S_{abcd}M_{ijd}.
\end{align*}
for which we have conveniently defined
\[
A_{abci}=-2K_{ic[a;b]},\qquad A_{abij}=-2M_{ij[a;b]}-K_{aic}K_{jbc}+K_{ibc}K_{jac}.
\]
i.e., the structural equation for $\omega_{ab}$ is now written
\[
d\omega_{ab}=-\omega_{ac}\wedge\omega_{cb}+\tfrac{1}{2}S_{abcd}\,\omega_{c}\wedge\omega_{d}+A_{abci}\,\omega_{c}\wedge\pi_{i}+\tfrac{1}{2}A_{abij}\,\pi_{i}\wedge\pi_{j}.
\]

Next we come to the generic relations. Since all our invariants are tensors, the generic relations are simply the commutation properties of our covariant derivatives.
 Let us try an examples. Let $I$ be a scalar quantity. For its second order covariant derivatives, any algebraic relations are obtained by calculating $d^{2}I$. We have
\[
d^{2}I=I_{;ab}\,\omega_{a}\wedge\omega_{b}+(I_{;ai}-I_{;ia}+K_{iab}I_{;b})\omega_{i}\wedge\omega_{a}+(I_{;jk}-I_{;a}M_{jka})\omega_{k}\wedge\omega_{j}.
\]
So we have
\begin{align*}
  I_{;ab}-I_{;ba}&=0,\\
  I_{;ai}-I_{;ia}&=-K_{iab}I_{;b},\\
  I_{;kl}-I_{;kj}&=I_{;a}M_{jka}.
\end{align*}
The non-zero right hand sides show non-commutativity. However, note that the right hand side contains only derivatives of order $1$ or less.

The same holds for tensor quantities. The general rule for exchanging orders of derivations is complicated, but it can be seen from the following example:
\begin{align*}
  d^{2}T_{ia}{}={}&(T_{ia;kj}-T_{ia;b}M_{jkb}-\tfrac{1}{2}T_{la}S_{iljk}-\tfrac{1}{2}T_{ib}A_{abjk})\omega_{j}\wedge\omega_{k}\\
  &+(T_{ia;dc}-\tfrac{1}{2}T_{ib}S_{abcd})\omega_{c}\wedge\omega_{d}\\
  &+(T_{ia;bj}-T_{ia;jb}+T_{ia;c}K_{jc;b}+T_{ic}A_{acbj})\omega_{j}\wedge\omega_{b}.
\end{align*}
What is important from these relations is that, \emph{if we ignore derivatives of lower order}, then covariant derivatives commute.

There now only remains what we called the derived relations. Again, since all our invariants, and also their covariant derivatives, are tensors, the derived relations are obviously the following: for example, if we have a relation $P_{ijab}=0$, then we have also $P_{ijab;l}=0$, $P_{ijab;c}=0$, $P_{ijab;lm}=0$, $P_{ijab;cd}=0$, $P_{ijab;lc}=0$, $P_{ijab;cl}=0$, etc. If we have a product of two invariants, for example, $A_{a}B_{i}=0$, then by considering the exterior derivative of the appropriate relations, it is easy to see that the ``Leibniz rule'' applies, and we have $A_{a;j}B_{i}+A_{a}B_{i;j}=0$, $A_{a;b}B_{i}+A_{a}B_{i;b}=0$, etc.

\subsection{The degrees of freedom of a Riemannian submersion}
We have essentially formulated a Riemannian submersion in terms of moving frames, and obtained all the algebraic relations among the differential invariants of the system. As discussed in \cite{hu1}, in this case we can deduce the \emph{degree of freedom} of this system with only minimal efforts. Let us now do this, by applying the algorithm proposed in \cite{hu1}.

The first thing we need to do is to find, among the differential invariants, an algebraically independent set whose indices are arranged in a suitable (preferably decreasing) order. Obviously, the following can be without doubt taken as independent:
\begin{center}
  \begin{tabular}{cc}
    Invariant&Independent terms\\
    \hline
    $M_{ija}$&$i>j$\\
    $K_{iab}$&$a\ge b$\\
    $S_{ijkl}$&$i>j,\ k>l,\ i\ge k,\ j\ge l$\\
    $S_{abcd}$&$a>b,\ c>d,\ a\ge c,\ b\ge d$\\
    $S_{ijkl;m}$&$i>j,\ k>l,\ i\ge k,\ j\ge l,\ k\ge m$\\
    $S_{abcd;e}$&$a>b,\ c>d,\ a\ge c,\ b\ge d,\ c\ge e$
  \end{tabular}
\end{center}
the arrangement indices on the Riemann tensors comes directly from the discussion of Riemannian geometry in \cite{hu1}.

In addition to the above, the first Bianchi identities give us two relations
\[
M_{[ij|a;|k]}=\cdots,\qquad M_{ij(a;b)}=-K_{[i|ab;|j]},
\]
using which, we can set
\begin{center}
  \begin{tabular}{cc}
    Invariant&Independent terms\\
    \hline
    $M_{ij[a;b]}$&$i>j,\ a>b$\\
    $M_{ija;k}$&$i>j,\ i\ge k$\\
    $K_{iab;c}$&$a\ge b$\\
    $K_{iab;j}$&$a\ge b$\\
  \end{tabular}
\end{center}
The only one that may require some explanation is the second one. Indeed, for $M_{ija;k}$, consider the indices $i,j,k$ to be all distinct. To be concrete, we can write them as $1$, $2$, $3$. Then we can list all quantities with these indices:
\[
M_{12a;3},\quad
M_{23a;1},\quad
M_{31a;2},\quad
M_{32a;1},\quad
M_{21a;3},\quad
M_{13a;2}.
\]
Using $M_{(ij)a;b}=0$, all those with $i<j$ can be expressed in terms of those with $i>j$. Hence we are left with
\[
M_{31a;2},\quad
M_{32a;1},\quad
M_{21a;3}.
\]
There is exactly one relation among these three quantities:
\[
M_{31a;2}-M_{21a;3}-M_{32a;1}=\text{ functions of zeroth order invariants},
\]
so we can express $M_{21a;3}$ in terms of the other two and zeroth order invariants. Hence in this case we can take all normal expressions to satisfy $i>j$ and $i>k$. 

Consider the case where there are only two distinct indices, and to be concrete let us assume that they are $1$ and $2$. Then we have the terms
\[
M_{12a;1},\quad
M_{12a;2},\quad
M_{21a;1},\quad
M_{21a;2},\quad
\]
Using the antisymmetry in the first two indices, we can reduce this set to
\[
M_{21a;1},\qquad M_{21a;2}.
\]
In this case the relation $M_{[ij|a;|k]}$ is satisfied identically. Since $M_{(ij)a}=0$, we cannot have all indices $i$, $j$, $k$ identical. Hence, we see easily that in all these cases, the quantities are normal if and only if
\[
i>j,\qquad i\ge k.
\]

The second Bianchi identities, together with the generic and derived relations, gives us the symmetries for all the remaining invariants. For the second Bianchi identities, the ``interesting'' ones (i.e., the ones that are not of the form of a symmetry of a Riemannian tensor) are
\begin{align*}
  S_{abcd;i}&=2A_{abci;d}+\cdots,\\
  A_{ab[ij;k]}&=\cdots,\\
  A_{abc[i;j]}&=-\tfrac{1}{2}A_{abij;c}+\cdots.
\end{align*}
where dots denote terms of lower order. Expressing $A_{abij}$ and $A_{abci}$ in terms of the derivatives of $M_{ija}$ and $K_{iab}$, these become
\begin{align*}
  S_{abcd;i}&=-2K_{ic[a;b]d}+\cdots,\\
  M_{[ij|[ab];|k]}&=\cdots,\\
  K_{ic[a;b]j}&=M_{ij[a;b]c}+\cdots.
\end{align*}
The ghastly notation $M_{[ij|[ab];|k]}$ simply means
\[
\tfrac{1}{3}(M_{ij[ab];k}+M_{jk[ab];i}+M_{ki[ab];c}).
\]
Hence for the remaining invariants: $S_{ijkl;a}$ all vanish, $S_{abcd;i}$ we take to be independent. There remains
\[  M_{ij[a;b]c},\qquad M_{ij[a;b]k}, \qquad M_{ija;kl},\qquad
  K_{iab;cd},\qquad K_{iab;cj},\qquad K_{iab;jk}.
\]
$M_{ija;kl}$ will have normal terms satisfying
\[
i>j,\qquad i\ge k,\qquad k\ge l.
\]
Using the relation for $M_{[ij|[a;b]|k]}$, $M_{ij[a;b]k}$ will have normal terms satisfying
\[
a>b,\qquad i>j,\qquad i\ge k.
\]
For $M_{ij[a;b]c}$, it contains no normal terms since by our index preference they are expressed in terms of $K_{ic[a;b]j}$.

$K_{iab;jk}$ will have normal terms
\[
a\ge b,\qquad j\ge k,
\]
where as $K_{iab;cj}$ simply has
\[
a\ge b.
\]
The most important term is $K_{ica;bd}$. First of all, $K_{ic[a;b]d}$ is not independent since it is expressible in terms of $S_{abcd;i}$. Hence we should only consider $K_{ic(a;b)d}$. For $K_{ica;bd}$, we can swap the first two or last two indices. For the middle two, we have
\[
K_{ica;bd}=K_{icb;ad}-S_{abcd;i}+\cdots
\]
hence for counting purposes, \emph{these four indices are totally symmetric.} We can arrange $K_{iab;cd}$ such that
\[
a\ge b\ge c\ge d.
\]
So finally, we have a table of independent invariants

\begin{center}
  \begin{tabular}{cc}
    Invariant&Independent terms\\
    \hline
    $M_{ija}$&$i>j$\\
    $K_{iab}$&$a\ge b$\\
    $S_{ijkl}$&$i>j,\ k>l,\ i\ge k,\ j\ge l$\\
    $S_{abcd}$&$a>b,\ c>d,\ a\ge c,\ b\ge d$\\
    $M_{ija;b}$&$i>j,\ a>b$\\
    $M_{ija;k}$&$i>j,\ i\ge k$\\
    $K_{iab;c}$&$a\ge b$\\
    $K_{iab;j}$&$a\ge b$\\
    \hline
    $S_{abcd;i}$&$a>b,\ c>d,\ a\ge c,\ b\ge d$\\
    $S_{ijkl;m}$&$i>j,\ k>l,\ i\ge k,\ j\ge l,\ k\ge m$\\
    $S_{abcd;e}$&$a>b,\ c>d,\ a\ge c,\ b\ge d,\ c\ge e$\\
    $M_{ija;kl}$&$i>j,\ i\ge k,\ k\ge l$\\
    $M_{ija;bk}$&$a>b,\ i>j,\ i\ge k$\\
    $K_{iab;jk}$&$a\ge b,\ j\ge k$\\
    $K_{iab;cj}$&$a\ge b$\\
    $K_{iab;cd}$&$a\ge b\ge c\ge d$
  \end{tabular}
\end{center}
which includes all derived invariants of the Riemannian tensors $S_{ijkl}$ and $S_{abcd}$ up to first order, and all derived invariants of the gluing tensors $M_{ija}$ and $K_{iab}$ up to second order. 

We can take the second block above to be the involutive seeds: this means that for our algorithm, we take the exterior differential system to be
\begin{equation}
  \label{eq:diffset1}
  \left\{
    \begin{aligned}
      dM_{ija}&=M_{ija;k}\omega_{k}+M_{ija;b}\omega_{b}+\cdots,\\
      dK_{iab}&=K_{iab;j}\omega_{j}+K_{iab;c}\omega_{c}+\cdots,
    \end{aligned}
  \right.
\end{equation}
together with
\begin{equation}
  \label{eq:diffset2}
  \left\{
    \begin{aligned}
      dM_{ija;k}&=M_{ija;kl}\omega_{l}+M_{ija;bk}\omega_{b}+\cdots,\\
      dM_{ija;b}&=M_{ija;bk}\omega_{k}+M_{ija;bc}\omega_{c}+\cdots,\\
      dK_{iab;j}&=K_{iab;jk}\omega_{k}+K_{iab;cj}\omega_{c}+\cdots,\\
      dK_{iab;c}&=K_{iab;cj}\omega_{j}+K_{iab;cd}\omega_{d}+\cdots,\\
      dS_{ijkl}&=S_{ijkl;m}\omega_{m}+\cdots,\\
      dS_{abcd}&=S_{abcd;i}\omega_{i}+S_{abcd;e}\omega_{e}+\cdots.
    \end{aligned}
  \right.
\end{equation}
The exterior derivation of \eqref{eq:diffset1} vanishes identically if we use \eqref{eq:diffset2}. The \emph{algebraically independent} quantities that are written explicitly on the right hand side of \eqref{eq:diffset2} are the involutive \emph{seeds}.

It can be verified that the conditions of involutive ordering are satisfied if we take all indices $a,b,c\dots$ to be greater than all indices $i,j,k\dots$. Thus, from the algorithm, the number of degree of freedom is given by the number of seeds whose last index is maximal. These are obtained only from
\[
S_{abcd;e}\qquad\text{for}\qquad a=c=e=q, \ d\le b<q
\]
and
\[
K_{iab;cd}\qquad\text{for}\qquad a=b=c=d=q.
\]
These give a total of 
\[
s_{p+q}=\frac{q(q-1)}{2}+p
\]
functions of $(p+q)$ variables. 

There are two other characters that may be of interest:
\begin{align*}
  s_{p+1}&=\frac{q^{2}(q^{2}-1)}{2}+\frac{pq^{2}(q+1)}{2},\\
  s_{p}&=\frac{p(p-1)}{2}+\frac{q(q+1)[q^{2}-q-1+p(q+2)]}{2}.
\end{align*}
If $p=0$, $s_{p+q}$ gives the degree of freedom for $q$ dimensional Riemannian space. If $q=0$, $s_{p}$ gives the degree of freedom for $p$ dimensional Riemannian space (in this case the formula for $s_{p+q}$ does not make sense). The character $s_{p+1}$ gives the minimal number of equations we need so as to kill all degrees of freedom on the leaves. Caveat: this counting includes all derived equations up to the order we are considering, for example, if we specify $M_{ija}=0$, we automatically have also $M_{ija;k}=0$, $M_{ija;b}=0$, etc. Needless to say, these two characters, being non-maximal characters in the general case, depend on to which order to which we take the differential invariants: here first order for the Riemann tensors, and second order for the gluing invariants.

\subsection{Existence of Riemannian submersions}
\label{sec:exist-struct-pres}
We now come the following problem: given a Riemannian geometry, does there exist a structural preserving submersion on it? A first attempt would be to use \eqref{eq:oldriemcurv} and carry out the involutive procedure, where the right hand sides are now taken to be given functions. This is, however, extremely messy.
Instead, in \cite{hu1} it was shown that the degree of freedom of a general $p+q$ dimensional Riemannian geometry is
\[
s_{p+q}=\frac{(p+q)(p+q-1)}{2}.
\]
The difference of this degree of freedom and the one we have found for a general structure preserving Riemannian submersion is
\[
\frac{p(p+2q-3)}{2}.
\]
This number is greater than zero except for the case of $p=q=1$ (if $p$ or $q$ is zero, then the submersion is trivial). Thus, {except for the case of $p=q=1$, Riemannian spaces that admit structure-preserving submersions are \emph{exceptional}} (they have measure zero in the space of all Riemannian geometries, roughly speaking).

For $p=q=1$, the two sets of degrees of freedom match, so it could be that all $2$ dimensional Riemannian spaces admit structural preserving submersions. We will now prove that in the analytic case, this possibility is locally realised (globally, there might be topological obstructions).

Indeed, in two dimensions, the structural equation for a Riemannian submersion is exceptionally simple:
\[
\left\{
  \begin{aligned}
    d\omega_{0}&=K\omega_{1}\wedge\omega_{0},\\
    d\omega_{1}&=0,
  \end{aligned}
\right.
\]
where $\omega_{0}$ lives on the leaf, $\omega_{1}$ lives on the base, and there is no principal bundle: the reduction of the principal bundle of $SO(2)$ is complete. In other words, as long as we can choose a section of the bundle of $2$ dimensional Riemannian geometry such that the structural equation takes the above form, this section, with its distinguished directions $\omega_{0}$ and $\omega_{1}$, furnishes a Riemannian submersion.

A general section gives
\[
\left\{
  \begin{aligned}
    d\theta_{0}&=a\theta_{1}\wedge\theta_{0},\\
    d\theta_{1}&=b\theta_{1}\wedge\theta_{0}.
  \end{aligned}
\right.
\]
We want to find a function $t$ of two variables such that
\[
d(\cos t\,\theta_{0}+\sin t\,\theta_{1})=0,
\]
then we can set $\omega_{0}=\cos t\,\theta_{0}+\sin t\,\theta_{1}$, and we are done. Expanding the above, we get
\[
(-t_{,1}\sin t-t_{,0}\cos t+a+b)\theta_{1}\wedge\theta_{0}=0.
\]
Since now both $d\theta_{0}=0\pmod{\theta_{0}}$ and $d\theta_{1}=0\pmod{\theta_{1}}$, we can set $\theta_{0}=dx$, $\theta_{1}=dy$ for a certain system of coordinates $(x,y)$. Then the equation in question becomes
\[
\sin t\frac{\pd t}{\pd y}+\cos t\frac{\pd t}{\pd x}=a(x,y)+b(x,y),
\]
and this system is of Cauchy-Kowalewski form, hence provided $a(x,y)$ and $b(x,y)$ are analytic functions, solution always exists.

\subsection{The Cauchy data for Riemannian submersions.}
\label{sec:cauchy-data-riem}
The degree of freedom of the submersion we have calculated gives us the number of functions we need to specify to have a well defined Cauchy problem. However, taken at face value, it requires us to specify $K_{iab;cd}$ for $a=b=c=d=q$ and $S_{abcd;e}$ for $a=c=e=q$, $d\le b<q$, and these data are neither convenient nor very invariant. We will now propose some better ways of specifying the Cauchy data, which yields a well-defined Cauchy problem.

Our aim is to kill the above two terms in the list of involutive seeds. For $S_{abcd;e}$, we know what to do: by our discussion of Riemannian geometry, it suffices to specify the Ricci tensor. We need to check that when we lower the index $S_{abcd;e}$ when $e=1$ to $S_{abcd;i}$, we maintain independence: this does hold. For $K_{iab;cd}$, we need $q$ equations, and we will simply specify the contraction of $K_{iab}$:
\[
K_{i}\equiv\sum_{a}K_{iaa}.
\]
Hence for $K_{iab}$, we now take the invariant terms to be those where not both $a$ and $b$ take the maximal value. At first order, $K_{iab;c}$ and $K_{iab;j}$ also cannot have both $a$ and $b$ taking maximal value. At second order, we have an equation of the form
\[
S_{abcd;i}=K_{icb;ad}-K_{ica;bd}+\dots=2K_{ic[b;a]d}+\cdots.
\]
If the indices $a$, $b$ both take maximal value, this is an identity. If $a$ and $c$ take maximal value, then $K_{icb;ad}$ is no longer considered independent. This means for $K_{iab;cd}$ to be independent, when $a$ is maximal, we require $q=a>b\ge c\ge d$. There is also no problem when we lower $K_{iab;cd}$ to $K_{iab;cj}$.

Granted these, the character $s_{p+q}$ is zero now. The contribution to $s_{p+q-1}$ now comes from two parts. The first part, having its origin in $S_{abcd;e}$ on which the Ricci tensor condition has been imposed, is
\[
q(q-3)
\]
when $q\ge 3$. When $q=2$ it is zero, and for $q=1$ the case needs to be treated separately, since we pass directly to the reduced space (this case will be treated in a later chapter). There are only two contributions from $K_{iab;cd}$ now, namely
\[
K_{iab;cd},\qquad a=q\text{ or }q-1,\quad b=c=d=q-1.
\]
So the degree of freedom is now
\[
s_{p+q-1}=2p+(s_{q-1} \text{ for the Einstein theory of dimension }q).
\]

Instead of specifying the Ricci tensor, we can also directly specify the metric of the fibre at each point of the space. Then $S_{abcd}$ and all its derivatives are no longer independent, and it is easy to see that for this case, the degree of freedom is simply ($q>1$)
\[
s_{p+q-1}=2p,
\]
the system is still involutive, showing that \emph{it is always consistent to specify any geometry of the fibres independently at each point on the reduced manifold.} On the other hand, attempting to reduce the order where the first non-vanishing character occur down to the reduced manifold by directly specifying an equation on $K_{iab}$ would lead to compatibility problems, shown by the fact that under such constraints the system of invariants can no longer be considered a system of involutive seeds. 

\section{Weyl submersion of codimension $1$}

Here we study a special case of affine submersion: the Weyl submersion. The aim is to give a non-Riemannian illustration of our procedure, but more importantly, to derive results that will be used in the second part of this paper. As many calculations and reasoning are similar to the Riemannian case, we will be much more brief here.

\subsection{The structure of Weyl geometry as a generalised space.}
\label{sec:struct-conf-geom}
Weyl geometry is constructed by adding a scaling degree of freedom to Riemannian geometry \footnote{Due to the history of the discovery of gauge theories, this is often compared with the $U(1)$ principal bundle, i.e., electromagnetism. It is important to note that there are a few differences here and there. In particular, the electromagnetic connection is separate from the spacetime connection.} and, unlike the more general conformal geometry, remains an affine geometry. On the base manifold, we set up an orthonormal frame $\omega^{\mu}$ (the ``normal'' part now only makes limited sense: it is no longer possible to compare the length of two covectors not situated in the same cotangent space). As the local symmetry group is now larger than the rotational group, from now on we need to pay attention if an index is upstairs or downstairs. On the bundle, which is now $M\times SO(n)\times\rs^{+}$, the coframe is formed by $\omega^{\mu}$, $\omega^{\mu}{}_{\nu}$ and $\omega^{\lambda}{}_{\bar\lambda}\equiv \tau$, with the structural equation
\[
\left\{
  \begin{aligned}
    d\omega^{\mu}&=-\omega^{\mu}{}_{\nu}\wedge\omega^{\nu}-\tau\wedge\omega^{\mu},\\
    d\omega^{\mu}{}_{\nu}&=-\omega^{\mu}{}_{\lambda}\wedge\omega^{\lambda}{}_{\nu}+\tfrac{1}{2}R^{\mu}{}_{\nu\rho\lambda}\omega^{\rho}\wedge\omega^{\lambda},\\
    d\tau&=\tfrac{1}{2}F_{\mu\nu}\omega^{\mu}\wedge\omega^{\nu},
  \end{aligned}
\right.
\]
where $F_{\mu\nu}$ is the \emph{scaling curvature}.

The \emph{covariant derivative} for any affine theory, acting on tensor and form components, are defined by
\[
dv^{i}=v^{i}{}_{;j}\omega^{j}-v^{k}\omega^{i}{}_{k},\qquad dw_{i}=w_{i;j}\omega^{j}+w_{k}\omega^{k}{}_{i}
\]
so for our present case
\[
dv^{\mu}=v^{\mu}{}_{;\nu}\omega^{\nu}-v^{\lambda}\omega^{\mu}{}_{\lambda}-v^{\mu}\tau,\qquad dw_{\mu}=w_{\mu;\nu}\omega^{\nu}+w_{\lambda}\omega^{\lambda}{}_{\mu}+w_{\mu}\tau,
\]
the placement of indices dictates whether we get a plus or a minus term linear in $\tau$, the scaling connection. 

By a reasoning entirely analogous to the Riemannian case, we can show that all invariants are again tensors. Thus, 
\begin{align*}
  dF_{\mu\nu}&=F_{\mu\nu;\lambda}\omega^{\lambda}+F_{\lambda\nu}\omega^{\lambda}{}_{\mu}+F_{\mu\lambda}\omega^{\lambda}{}_{\nu}+2F_{\mu\nu}\tau,\\
  dR^{\mu}{}_{\nu\rho\lambda}&=R^{\mu}{}_{\nu\rho\lambda;\gamma}\omega^{\gamma}-R^{\gamma}{}_{\nu\rho\lambda}\omega^{\mu}{}_{\gamma}+R^{\mu}{}_{\gamma\rho\lambda}\omega^{\gamma}{}_{\nu}+R^{\mu}{}_{\nu\gamma\lambda}\omega^{\gamma}{}_{\rho}+R^{\mu}{}_{\nu\rho\gamma}\omega^{\gamma}{}_{\lambda}+2R^{\mu}{}_{\nu\rho\lambda}\tau.
\end{align*}
Besides the ``defining'' symmetries,
\[
F_{\mu\nu}=-F_{\nu\mu},\qquad R^{\mu}{}_{\nu\rho\lambda}=-R^{\nu}{}_{\mu\rho\lambda}=-R^{\mu}{}_{\nu\lambda\rho},
\]
we have the Bianchi identities:
\[
\left\{
  \begin{aligned}
    &d^{2}\omega^{\mu}:&R^{\mu}{}_{[\nu\rho\lambda]}&=-\delta^{\mu}{}_{[\rho}F_{\nu\lambda]},\\
    &d^{2}\tau:&F_{[\mu\nu;\rho]}&=0,\\
    &d^{2}\omega^{\mu}{}_{\nu}:&R^{\mu}{}_{\nu[\rho\lambda;\gamma]}&=0.
  \end{aligned}
\right.
\]
We have the table of involutive seeds
\begin{center}
  \begin{tabular}{cc}
    Invariant&Normal terms\\
    \hline
    $F_{\mu\nu}$&$\mu>\nu$\\
    $R^{\mu}{}_{\nu\rho\lambda}$&Riemann tensor symmetry\\
    \hline
    $F_{\mu\nu;\lambda}$&$\mu>\nu,\ \mu\ge \lambda$\\
    $R^{\mu}{}_{\nu\rho\lambda;\gamma}$&Riemann tensor symmetry
  \end{tabular}
\end{center}
so the degree of freedom is
\[
s_{n}=\frac{n(n-1)}{2}+(n-1)=\frac{(n+2)(n-1)}{2}.
\]

We also know that the maximal number of symmetries of the space is equal to the dimension of the symmetry group of the homogeneous version of the space, which is in the present case $n+\dim(SO(n))+1$, corresponding to translation, rotation and scaling. For example, in Cartesian coordinates with the Euclidean metric, the ``Killing'' vector fields are
\[
\frac{\pd }{\pd x^{i}},\qquad x^{i}\frac{\pd}{\pd x^{j}}-x^{j}\frac{\pd}{\pd x^{i}},\qquad \sum_{i}x^{i}\frac{\pd}{\pd x^{i}}.
\]

\subsection{The structural equations of Weyl submersion of codimension $1$.}
\label{sec:struct-pres-subm}

Now we can construct the theory of structure preserving submersion in Weyl geometry. For simplicity we shall restrict the codimension one case. As in the Riemannian case, let the horizontal forms on the total space to be divided into two classes, $\omega^{0}$ and $\omega^{i}$, and let the horizontal forms on the reduced space be $\pi^{i}$. On the product space we require
\[
\omega^{i}=\pi^{i}.
\]
The reduction of the principal bundle entails, as usual
\[
\omega^{0}{}_{i}=-\omega^{i}{}_{0}=K_{i}{}^{0}{}_{0}\omega^{0}-M_{ij}{}^{0}\omega^{j}-B_{ij}{}^{0}\omega^{j},
\]
where
\[
M_{ij}{}^{0}=-M_{ij}{}^{0},\qquad B_{ij}{}^{0}=B_{ji}{}^{0}.
\]
and the first structural equations of the total space are, after reduction
\[
\left\{
  \begin{aligned}
    d\omega^{0}&=-K_{i}{}^{0}{}_{0}\omega^{0}\wedge\omega^{i}+M_{ij}{}^{0}\omega^{j}\wedge\omega^{i}-\tau\wedge\omega^{0},\\
    d\omega^{i}&=-\omega^{i}{}_{j}\wedge\omega^{j}-M^{i}{}_{j0}\omega^{j}\wedge\omega^{0}-B^{i}{}_{j0}\omega^{j}\wedge\omega^{0}-\tau\wedge\omega^{i},
  \end{aligned}
\right.
\]
we explicitly indicate all indices, including $0$, since now in general $M_{ij0}\neq M_{ij}{}^{0}$.

For the reduced space, the first structural equation is
\[
d\pi^{i}=-\pi^{i}{}_{j}\wedge\pi^{j}-\varpi\wedge\pi^{i}.
\]
Requiring $d\omega^{i}=d\pi^{i}$ now, we get
\[
(\pi^{i}{}_{j}-\omega^{i}{}_{j}+M^{i}{}_{j0}\omega^{0})+(B^{i}{}_{j0}\omega^{0}+\delta^{i}{}_{j}\varpi-\delta^{i}{}_{j}\tau)=C^{i}{}_{jk}\omega^{k},
\]
where $C^{i}{}_{jk}=C^{i}{}_{kj}$. First let us antisymmetrise the indices $i$, $j$ in this equation. This gives us
\[
\omega^{i}{}_{j}=\pi^{i}{}_{j}+M^{i}{}_{j0}\omega^{0}
\]
as usual. For the symmetric part, if $i$ and $j$ are distinct, we have
\[
B^{i}{}_{j0}\omega^{0}=C^{i}{}_{jk}\omega^{k},
\]
and since $\omega^{0}$ and $\omega^{i}$ are independent, both sides vanish. In particular, this shows that the only components of $C^{i}{}_{jk}$ that may be non-zero are those that have all three indices the same. If $i$ and $j$ are the same, then
\[
B^{i}{}_{\bar i0}\omega^{0}+\varpi-\tau=C^{i}{}_{\bar i\tilde i}\omega^{\vec i}.
\]
This must hold for all choice of indices $i$, hence 
\[
C^{i}{}_{jk}=0,\qquad B^{i}{}_{\bar i0}\equiv E_{0},\qquad B^{i}{}_{j0}=\delta^{i}{}_{j}E_{0}
\]
and
\[
\tau=\varpi+E_{0}\omega^{0}.
\]

Using the quantities $M_{ij}{}^{0}$ and $E_{0}$, we can exchange the forms $\pi^{i}{}_{j}$ and $\varpi$ for $\omega^{i}{}_{j}$ and $\tau$. Henceforth we take $\omega^{i}$, $\omega^{0}$, $\pi^{i}{}_{j}$, $\tau$ to be the coframe (connection) on the total space, so that the coframe derivative (covariant derivative) in the $i$ direction is independent of the fibre coordinates.

Now the complete structural equations for the coframe are
\[\left\{
  \begin{aligned}    
    d\omega^{0}&=-K_{i}{}^{0}{}_{0}\omega^{0}\wedge\omega^{i}-M_{ij}{}^{0}\omega^{i}\wedge\omega^{j}-\varpi\wedge\omega^{0},\\
    d\omega^{i}&=-\pi^{i}{}_{j}\wedge\omega^{j}-\varpi\wedge \omega^{i},\\
    d\pi^{i}{}_{j}&=-\pi^{i}{}_{k}\wedge\pi^{k}{}_{j}+\tfrac{1}{2}S^{i}{}_{jkl}\omega^{j}\wedge\omega^{l},\\
    d\varpi&=\tfrac{1}{2}G_{ij}\omega^{i}\wedge\omega^{j}.
  \end{aligned}
\right.
\]
The last three equations are equations on the reduced space.

\subsection{Algebraic relations and the degree of freedom}
\label{sec:bianchi-identities}

It is now customary to derive the Bianchi identities for the submersion. First,
\[
d^{2}\omega^{0}=-(\tfrac{1}{2}G_{ij}-K_{i}{}^{0}{}_{0;j}-M_{ij}{}^{0}{}_{;0})\omega^{i}\wedge\omega^{j}\wedge\omega^{0}-(M_{ij}{}^{0}{}_{;k}-K_{i}{}^{0}{}_{0}M_{jk}{}^{0})\omega^{i}\wedge\omega^{j}\wedge\omega^{k},
\]
so
\begin{equation}
  \label{eq:mb9}
G_{ij}=-2K_{[i|}{}^{0}{}_{0|;j]}-2M_{ij}{}^{0}{}_{;0},\qquad M_{[ij|}{}^{0}{}_{;|k]}=K_{[i|}{}^{0}{}_{0}M_{|jk]}{}^{0}.  
\end{equation}
Next,
\[
d^{2}\varpi=\tfrac{1}{2}G_{ij;k}\omega^{i}\wedge\omega^{j}\wedge\omega^{k}+\tfrac{1}{2}G_{ij;0}\omega^{i}\wedge\omega^{j}\wedge\omega^{0},
\]
giving us
\[
G_{[ij;k]}=0,\qquad G_{ij;0}=0.
\]
In particular, this shows that $G_{ij}$ is independent of the fibre coordinates, which should be expected.

Next,
\[
d^{2}\omega^{i}=-\tfrac{1}{2}(S^{i}{}_{jkl}-\delta^{i}{}_{l}G_{jk})\omega^{j}\wedge\omega^{k}\wedge\omega^{l},
\]
so
\[
S^{i}{}_{[jkl]}=\delta^{i}{}_{[l}G_{jk]},
\]
which is the extension of the first Bianchi identity. We can get a clearer picture of how these two quantities are related by forming the contraction of the Ricci tensor:
\begin{equation}
  \label{eq:imaginebreaker}
S_{[jl]}=\tfrac{p-2}{2}G_{jl}  
\end{equation}
where $p$ is the dimension of the reduced space, or $n-1$ in our case. Hence the scaling curvature contributes to the antisymmetric part of the Ricci tensor. Finally,
\[
d^{2}\pi^{i}{}_{j}=\tfrac{1}{2}S^{i}{}_{jkl;m}\omega^{m}\wedge\omega^{k}\wedge\omega^{l}+\tfrac{1}{2}S^{i}{}_{jkl;0}\omega^{0}\wedge\omega^{k}\wedge\omega^{l},
\]
which gives
\[
S^{i}{}_{j[kl;m]}=0,\qquad S^{i}{}_{jkl;0}=0,
\]
which are the usual second Bianchi identity and the condition that $S^{i}{}_{jkl}$ is independent of the fibre coordinates.

Armed with these identities, we can arrange the invariants
\begin{center}
  \begin{tabular}{cc}
    Invariant&Independent terms\\
    \hline
    $E_{0}$&all\\
    $M_{ij0}$&$i>j$\\
    $K_{i00}$&all\\
    $S_{ijkl}$&$i>j,\ k>l,\ i\ge k,\ j\ge l$\\
    $G_{ij}$&$i>j$\\
    $E_{0;i}$&all\\
    $E_{0;0}$&all\\
    $M_{ij0;k}$&$i>j,\ i\ge k$\\
    $K_{i00;0}$&all\\
    $K_{i00;j}$&all\\
    \hline
    $S_{ijkl;m}$&$i>j,\ k>l,\ i\ge k,\ j\ge l,\ k\ge m$\\
    $G_{ij;k}$&$i>j,\ i\ge k$\\
    $M_{ij0;kl}$&$i>j,\ i\ge k,\ k\ge l$\\
    $K_{i00;jk}$&$j\ge k$\\
    $K_{i00;0j}$&all\\
    $K_{i00;00}$&all\\
    $E_{0;ij}$&$i\ge j$\\
    $E_{0;0i}$&all\\
    $E_{0;00}$&all
  \end{tabular}
\end{center}
Again, the second block can be taken as involutive seeds.
Now in addition to $K_{i00;00}$, $E_{0;00}$ also contributes to the Cartan character. We see that, in the generic case, the problem of structure preserving submersion in Weyl geometry has
\[
s_{n}=n
\]
degree of freedom, one more than in the Riemannian case.

\subsection{Invariants of the total space}
\label{sec:invar-total-space}

We also want to get expressions of the curvatures $F_{\mu\nu}$ and $R^{\mu}{}_{\nu\rho\lambda}$ in terms of the curvatures $G_{ij}$, $S^{i}{}_{jkl}$ and the invariants $K_{i}{}^{0}{}_{0}$, $M_{ij}{}^{0}$ and $E_{0}$. We just have to calculate.

From
\[
d\tau=F_{i0}\omega^{i}\wedge\omega^{0}+\tfrac{1}{2}F_{ij}\omega^{i}\wedge\omega^{j}
=d(\varpi+E_{0}\omega^{0}),  
\]we get
\begin{equation}\label{eq:magickiller}
\left\{
  \begin{aligned}
    F_{ij}&=G_{ij}-2E_{0}M_{ij}{}^{0},\\
    F_{i0}&=E_{0;i}+E_{0}K_{i}{}^{0}{}_{0}.
  \end{aligned}
\right.  
\end{equation}
From
\[d\omega^{i}{}_{j}=-\omega^{i}{}_{k}\wedge\omega^{k}{}_{j}+\tfrac{1}{2}R^{i}{}_{jkl}\omega^{k}\wedge\omega^{l}+R^{i}{}_{jk0}\omega^{k}\wedge\omega^{0}-\omega^{i}{}_{0}\wedge\omega^{0}{}_{j}
=d(\pi^{i}{}_{j}+M^{i}{}_{j0}\omega^{0}),  
\]we get
\[
\left\{
  \begin{aligned}
    R^{i}{}_{jkl}{}={}&{}S^{i}{}_{jkl}-2M^{i}{}_{j0}M_{kl}{}^{0}+M^{i}{}_{l0}M_{jk}{}^{0}-M^{i}{}_{k0}M_{jl}{}^{0}\\
    &{}+M_{jk}{}^{0}\delta^{i}{}_{l}E_{0}-M_{jl}{}^{0}\delta^{i}{}_{k}E_{0}+M^{i}{}_{l}\delta_{jk}E^{0}-M^{i}{}_{k}\delta_{jl}E^{0}\\
    &{}+\delta^{i}{}_{l}\delta_{jk}E_{0}E_{0}-\delta^{i}{}_{k}\delta_{jl}E_{0}E_{0},\\
    R^{i}{}_{jk0}{}={}&M^{i}{}_{j0;k}-M_{jk}{}^{0}K^{i}{}_{00}+M^{i}{}_{k0}K_{j}{}^{0}{}_{0}+M^{i}{}_{j0}K_{k}{}^{0}{}_{0}\\
    &{}-\delta_{jk}E^{0}K^{i}{}_{00}+\delta^{i}{}_{k}E_{0}K_{j}{}^{0}{}_{0}.
  \end{aligned}
\right.
\]
From
\begin{align*}
d\omega^{i}{}_{0}&=-\omega^{i}{}_{j}\wedge\omega^{j}{}_{0}+\tfrac{1}{2}R^{i}{}_{0jk}\omega^{j}\wedge\omega^{k}+R^{i}{}_{0j0}\omega^{j}\wedge\omega^{0}\\
&=d(-K^{i}{}_{00}\omega^{0}+M^{i}{}_{j0}\omega^{j}+\delta^{i}{}_{j}E_{0}\omega^{j})  
\end{align*}
we get
\[
\left\{
  \begin{aligned}
    R^{i}{}_{0jk}{}={}&{}2K^{i}{}_{00}M_{jk}{}^{0}-M^{i}{}_{j0;k}+M^{i}{}_{k0;j}-\delta^{i}{}_{j}E_{0;k}+\delta^{i}{}_{k}E_{0;j},\\
    R^{i}{}_{0j0}{}={}&{}-M^{i}{}_{k0}M^{k}{}_{j0}-K^{i}{}_{00;j}-K^{i}{}_{00}K_{j}{}^{0}{}_{0}-M^{i}{}_{j0;0}\\
    &-\delta^{i}{}_{j}E_{0;0}-M^{i}{}_{k0}\delta^{k}{}_{j}E_{0}.
  \end{aligned}
\right.
\]

\section{Semi-Killing vector fields}

It should be noted that in all our exposition so far, we made no use of the concept of vector fields or Lie derivatives at all. As we have seen, this presented little problem, since differential forms and coframes are sufficient for us to derive the properties of structure-preserving submersions we want. More importantly, forms are better behaving objects than vectors, and in many geometries the concept of vectors is not very useful at all since it does not enjoy the invariance properties that we usually attribute with it in the Riemannian setting. Nonetheless, when the geometry is affine, this concept is useful, and since in physical applications we usually formulate problems starting from vectors instead of from forms, let us investigate how we can use vectors systematically in the study of submersions.

We will restrict our attention to the Riemannian case: in all affine cases the methods are similar.

\subsection{Killing vector fields}\index{isometries}
Let us first study Killing vectors. In a section of the principal bundle of a Riemannian geometry, the metric can be written as
\[
ds^{2}=\sum_{\mu}\omega_{\mu}\otimes\omega_{\mu}.
\]
Let $\mathbf{v}$ be a vector field on the manifold. If $\mathbf{v}$ is a Killing vector field, then
\begin{equation}
  \label{eq:killingvec}
  \mathcal{L}_{\mathbf{v}}\left(\sum\omega_{\mu}\otimes\omega_{\mu}\right)=0,
\end{equation}
where $\mathcal{L}$ denotes the Lie derivative.

However, this approach has the following disadvantage. We can calculate the Lie derivative using its Leibniz property and Cartan's ``magic formula''
\[
\mathcal{L}_{\mathbf{v}}\omega=\mathbf{v}\iprod(d\omega)+d(\mathbf{v}\iprod\omega),
\]
but $\mathbf{v}$ is a vector on the base manifold, and what should we do if we are confronted with something such as $\mathbf{v}\iprod\omega_{\mu\nu}$? Since we are now in a section of the principal bundle, we need to write $\omega_{\mu\nu}=a_{\mu\nu\lambda}\omega_{\lambda}$ and continue the calculation, but introducing additional variables $a_{\mu\nu\rho}$ in this way is undesirable. 

Let us reconsider our approach. The definition of the Killing vector is \eqref{eq:killingvec}, which involves the tensor product of the horizontal forms. Why not simply $\mathcal{L}_{\mathbf{v}}\omega_{\mu}=0$? Requiring that each horizontal form is separately invariant under the Killing vector is too strong: we only requires the invariance of the bilinear form. But it is obvious that for every Killing vector field that satisfies \eqref{eq:killingvec}, we can find a section of the principal bundle such that $\mathcal{L}_{\mathbf{v}}\omega_{\mu}=0$ holds: it suffices to use this relation as the definition of the section that we want.

But the relation $\mathcal{L}_{\mathbf{v}}\omega_{\mu}=0$, which is now required to hold in a particular section, has the following significance in the bundle: it is not only necessary to specify in which horizontal direction to move in order to obtain an isometry, but also to specify how we rotate the frames in this direction. In other words, let $\mathbf{V}$ be a vector field \emph{on the bundle} (in particular, it is \emph{not} assumed \emph{a priori} to be a tensor, and it has components $\mathbf{V}=V_{\mu}\mathbf{I}_{\mu}+{V}_{\mu\nu}\mathbf{I}_{\mu\nu}$, where $\mathbf{I}_{\mu}$ and $\mathbf{I}_{\mu\nu}$ are the dual basis for the coframe $\omega_{\mu}$, $\omega_{\mu\nu}$). The existence of an isometry requires
\[
\mathcal{L}_{\mathbf{V}}\omega_{\mu}=0
\]
on the bundle. Let us see what this implies by calculating the Lie derivative. We have
\begin{align*}
  \mathcal{L}_{\mathbf{V}}\omega_{\mu}&=-(V_{\rho}\mathbf{I}_{\rho}+V_{\rho\lambda}\mathbf{I}_{\rho\lambda})\iprod(\omega_{\mu\nu}\wedge\omega_{\nu})+d[(V_{\rho}\mathbf{I}_{\rho}+V_{\rho\lambda}\mathbf{I}_{\rho\lambda})\iprod\omega_{\mu}]\\
  &=-V_{\mu\nu}\omega_{\nu}+V_{\nu}\omega_{\mu\nu}+V_{\mu,\nu}\omega_{\nu}+V_{\mu,\rho\lambda}\omega_{\rho\lambda}
\end{align*}
where in the last line we have expanded $dV_{\mu}$ in terms of the coframe. We see that
\begin{equation}
  \label{eq:killingbundle}
  \left\{
  \begin{aligned}
    V_{\mu\nu}\omega_{\nu}&=V_{\mu,\nu}\omega_{\nu},\\
    V_{\nu}\omega_{\mu\nu}&=-V_{\mu,\rho\lambda}\omega_{\rho\lambda}.
  \end{aligned}
\right.
\end{equation}
The second equation is rather curious. Indeed, let $v_{\mu}$ be a vector on the base manifold, i.e., a tensor on the bundle. We know that
\[  dv_{\mu}\equiv v_{\mu;\nu}\omega_{\nu}+v_{\mu,\nu\rho}\omega_{\nu\rho}
  =v_{\mu;\nu}\omega_{\nu}-v_{\nu}\omega_{\mu\nu},
\]
giving us
\[
v_{\mu,\nu\rho}\omega_{\nu\rho}=-v_{\nu}\omega_{\mu\nu}.
\]
Hence the second equation of \eqref{eq:killingbundle} just tells us that $V_{\mu}$ are the components of a tensor, i.e., a vector on the base. Then the first equation gives us
\[
\left\{
  \begin{aligned}
    V_{[\mu;\nu]}&=V_{\mu\nu},\\
    V_{(\mu;\nu)}&=0.
  \end{aligned}
\right.
\]
The second equation gives us the usual Killing's equation in the bundle, whereas the first tells us how we need to \emph{lift} the Killing vector field on the base into a vector field on the bundle.

\subsection{Magic formula and derived relations}

In studies involving differential forms it is essential that we include the derived equations of equation. Even though when we use vector fields, strictly speaking we are not doing calculations with exterior differential systems, let us do the same nonetheless. Deriving the magic formula $\mathcal{L}_{\mathbf{v}}\omega=\mathbf{v}\iprod (d\omega)+d(\mathbf{v}\iprod\omega)$ gives
\[
d\mathcal{L}_{\mathbf{v}}\omega=\mathcal{L}_{\mathbf{v}}(d\omega),
\]
which is the well-known fact that exterior derivative and Lie derivatives commute on differential forms (deriving again yields an identity). Hence, \emph{when we have an equation concerning Lie derivatives, we should always include the derived equations as well}. Do this for $\mathcal{L}_{\mathbf{V}}\omega_{\mu}=0$, we have
\[
\mathcal{L}_{\mathbf{V}}(d\omega_{\mu})=-(\mathcal{L}_{\mathbf{V}}\omega_{\mu\nu})\wedge\omega_{\nu},
\]
so
\[
\mathcal{L}_{\mathbf{V}}\omega_{\mu\nu}=c_{\mu\nu\lambda}\omega_{\lambda},\qquad c_{\mu\nu\lambda}=c_{\mu\lambda\nu},
\]
but as $c_{\mu\nu\lambda}=-c_{\nu\mu\lambda}$, it vanishes identically. Hence even though we have only required $\mathcal{L}_{\mathbf{V}}\omega_{\mu}=0$, the vector field we have found on the bundle satisfies $\mathcal{L}_{\mathbf{V}}\omega_{\mu\nu}=0$ as well.

Now we derive the equation we have just obtained:
\[
\mathcal{L}_{\mathbf{V}}(d\omega_{\mu\nu})=\tfrac{1}{2}(\mathcal{L}_{\mathbf{V}}R_{\mu\nu\rho\lambda})\omega_{\rho}\wedge\omega_{\lambda},
\]
giving
\[
\mathcal{L}_{\mathbf{V}}R_{\mu\nu\rho\lambda}=0,
\]
since for the equation to be satisfied $R_{\mu\nu\rho\lambda}$ has to be both symmetric and antisymmetric in the $\rho$ and $\lambda$ indices.

Now we have obtained that the curvature \emph{tensor} $R_{\mu\nu\rho\lambda}$ is constant under the Killing vector field. We can derive these conditions further. For example, if $T_{\mu}$ is a tensor and $\mathcal{L}_{\mathbf{V}}T_{\mu}=0$, then
\[
\mathcal{L}_{\mathbf{V}}dT_{\mu}=\mathcal{L}_{V}(T_{\mu;\nu}\omega_{\nu}+\cdots)=(\mathcal{L}_{\mathbf{V}}T_{\mu;\nu})\omega_{\nu}=0,
\]
where dots indicate terms that are zero under Lie derivatives. Hence, by carrying out more derivations, we see that \emph{all covariant derivatives of the tensor are constant under the Lie derivative}. This result is easily seen to hold for tensors of all ranks.

\subsection{Semi-Killing vector fields and submersions}
\label{sec:anoth-view-struct}\index{Riemannian submersion}

We have seen that the condition of the existence of Killing vectors 
\begin{equation}
  \label{eq:kvs}  
\mathcal{L}_{\mathbf{V}}\omega_{\mu}=0
\end{equation}
 for all indices $\mu$ (which implies $\mathcal{L}_{\mathbf{V}}\omega_{\mu\nu}=0$) can be taken as the condition for isometry. Suppose that we have a Riemannian submersion. Then roughly speaking, the forms $\omega_{i}$, which are pullbacks of forms from the reduced manifold, are aligned along each leaf of the foliation (see {moving-frame-riem}). This suggests that we could try the condition $\mathcal{L}_{\mathbf{U}}\omega_{i}=0$, where $\mathbf{U}$ points only along the leaves on the manifold, as the condition for structure preserving submersion. Since \eqref{eq:kvs} now is only required to hold for some indices, we will call the vector field $\mathbf{U}$ a \emph{semi-Killing vector field}.

Since we have distinguished two subsets of horizontal forms $\omega_{i}$ and $\omega_{a}$, this amounts to a reduction of the principal bundle.
All group transformations that transforms \emph{between} these two sets are now forbidden and hence the corresponding to these group elements are no longer independent. Notice that this is \emph{different} from our reasoning using differential forms: indeed, such a reasoning relies on the property that the group action on the vectors, and hence on the forms, are \emph{linear}, and hence the group is affine. Effecting the reduction of the bundle, we can easily show that we must have
\[
\omega_{ai}=-\omega_{ia}=K_{iab}\omega_{b}-M_{ija}\omega_{j}
\]
as before by considering the allowed group transformations, but for the moment there is no constraints on $K_{iab}$ and $M_{ija}$. As for $\mathbf{U}$, we now have
\[
\mathbf{U}=U_{a}\mathbf{I}_{a}+U_{ij}\mathbf{I}_{ij}+U_{ab}\mathbf{I}_{ab},
\]
with $U_{a}\neq 0$ and no term in $\mathbf{I}_{i}$. We can now calculate
\begin{align*}
  \mathcal{L}_{\mathbf{U}}\omega_{i}&=d(\mathbf{U}\iprod\omega_{i})+\mathbf{U}\iprod d\omega_{i}\\
  &=(M_{ija}U_{a}-U_{ij})\omega_{j}+(K_{iab}-K_{iba})U_{b}\omega_{a},
\end{align*}
which requires
\[
M_{ija}U_{a}=U_{ij},\qquad K_{i[ab]}U_{b}=0.
\]
Now if there is only a single index for $a,b,\dots$, we obviously have
\begin{equation}
  \label{eq:conspsb}
  M_{(ij)a}=0,\qquad K_{i[ab]}=0,
\end{equation}
since we require $U_{a}\neq 0$.
If there are several indices for $a,b,\dots$, recall how these indices arise: these indices arise because we can do a submersion along $\mathbf{U}$. Hence there are as many $\mathbf{U}$ satisfying $\mathcal{L}_{\mathbf{U}}\omega_{i}=0$ as there are indices for $a,b,\dots$. Then we see that \eqref{eq:conspsb} is also satisfied in this case. On the other hand, if \eqref{eq:conspsb} is satisfied, obviously we can choose $U_{a}$ arbitrarily, then $U_{ij}$ is uniquely determined ($U_{ab}$ does not enter anywhere in these equations). But \eqref{eq:conspsb} is just the condition we found for structural preserving submersions using the method of integral varieties within product spaces. The two definitions are hence equivalent.

Note that now in general $\mathcal{L}_{\mathbf{U}}$ acting on $\omega_{a}$, $\omega_{ab}$ and $\omega_{ij}$ are not zero. But if we define
\[
\pi_{ij}=\omega_{ij}-M_{ija}\omega_{a},
\]
then
\[
\mathbf{U}\iprod\pi_{ij}=0
\]
and the derived equation of $\mathcal{L}_{\mathbf{U}}\omega_{i}$ gives
\[
\mathcal{L}_{\mathbf{U}}\pi_{ij}=0,
\]
we have recovered the modified connection we have defined. Carrying out the derivation further simply gives
\[
\mathcal{L}_{\mathbf{U}}S_{ijkl}=0,
\]
which should not surprise us now. We should not carry out further derivations: it is easier to carry out the analysis using the method of exterior differential systems as we have done before.

\subsection{Killing vector fields as semi-Killing vector fields}
\label{sec:isom-as-riem}

Now it is obvious from definition that every Killing vector field is a semi-Killing vector field. We will now derive what this implies if we view isometries as submersions.

To make calculation easier, first we change from the coframe with $\omega_{ij}$ to the coframe with $\varpi_{ij}$. The corresponding change for the frame is
\[
\mathbf{I}_{i}'=\mathbf{I}_{i},\qquad \mathbf{I}_{a}'=\mathbf{I}_{a}+M_{ija}\mathbf{I}_{ij},\qquad \mathbf{I}'_{ij}=\mathbf{I}_{ij},\qquad \mathbf{I}_{ab}'=\mathbf{I}_{ab},
\]
hence for a vector field $\mathbf{V}$ defining a direction of structural preserving submersion,
\[
\mathbf{V}=V_{a}\mathbf{I}'_{a}+V_{ab}\mathbf{I}_{ab}.
\]

Now suppose that $\mathbf{V}$ is also a Killing vector field. First recall a property of Lie derivatives: if $\mathcal{L}_{\mathbf{X}}\omega=0$ and $\mathcal{L}_{\mathbf{Y}}\omega=0$, then $\mathcal{L}_{a\mathbf{X}+b\mathbf{Y}}\omega=0$ where $a$ and $b$ are \emph{constants}. This means that if we have several Killing vector fields, we can form their linear combinations with constant coefficients which are still Killing vector fields. Next, the condition $\mathcal{L}_{\mathbf{V}}\omega_{\mu}=0$ is independent of whatever coframe we choose on the bundle. The easiest way to check this is to recall the properties enjoyed by the components of $\mathbf{V}$. Since $V_{\mu\nu}$ is obtained by lifting $V_{\mu}$, and this lifting needs to be done again once we change coframe, we only need to check the conditions on $V_{\mu}$. The condition that $V_{\mu}$ is a tensor is obviously invariant under change of frame, as well as $V_{(\mu;\nu)}=0$. Thus, for semi-Killing vector field, we only need to specify that it satisfy in addition $\mathcal{L}_{\mathbf{V}}\omega_{a}=0$, since $\mathcal{L}_{\mathbf{V}}\omega_{i}$ are already satisfied. 
Calculating,
\begin{align*}
  \mathcal{L}_{\mathbf{V}}\omega_{a}&=(V_{a,b}-V_{ab})\omega_{b}+(V_{a,i}-K_{iab}V_{b})\omega_{i}+V_{a,ij}\varpi_{ij}+(V_{a,bc}\omega_{bc}+V_{b}\omega_{ab})=0,
\end{align*}
giving the conditions
\[
\left\{
  \begin{aligned}
    V_{[a,b]}&=V_{ab},\\
    V_{(a,b)}&=0,\\
    V_{a,i}&=K_{iab}V_{b},\\
    V_{a,ij}&=0,\\
    V_{a,bc}&=V_{b}\delta_{ac}.
  \end{aligned}
\right.
\]
As usual, some of these conditions just mean that $V_{a}$ is the components of a tensor. Henceforth we will write
\[
dV_{a}=V_{a;b}\omega_{a}+K_{iab}V_{b}\omega_{i}-V_{b}\omega_{ab}.
\]

For the derived relations, in addition to the relations for general structure preserving submersions, we have
\[
\mathcal{L}_{\mathbf{V}}(d\omega_{a})=-(\mathcal{L}_{\mathbf{V}}\omega_{ab})\wedge\omega_{b}-(\mathcal{L}_{\mathbf{V}}K_{iab})\omega_{b}\wedge\omega_{i}-(\mathcal{L}_{\mathbf{V}}M_{ija})\omega_{i}\wedge\omega_{j}
\]
which immediately gives
\[
\mathcal{L}_{\mathbf{V}}K_{iab}=0,\qquad\mathcal{L}_{\mathbf{V}}M_{ija}=0,
\]
(the terms $\mathcal{L}_{\mathbf{V}}\omega_{ab}$ and $\mathcal{L}_{\mathbf{V}}K_{iab}\omega_{i}$ cannot mix, since the symmetries on the indices are opposite.)

 As for $\omega_{ab}$, 
\[
\mathcal{L}_{\mathbf{V}}\omega_{ab}=c_{abc}\omega_{c},
\]
but $c_{abc}$ has to be symmetric in $b$, $c$ but antisymmetric in $a$, $b$, so it vanishes identically. We have the additional condition
\[
\mathcal{L}_{\mathbf{V}}\omega_{ab}=0.
\]

Now for any tensor $T_{ab\dots ij\dots}$, if $\mathcal{L}_{\mathbf{V}}T_{ab\dots ij\dots}=0$, we have
\begin{align*}
  \mathcal{L}_{\mathbf{V}}(dT_{ab\dots ij\dots})&=\mathcal{L}_{\mathbf{V}}(T_{ab\dots ij\dots;c}\omega_{c}+T_{ab\dots ij\dots;k}\omega_{k}+\cdots)\\
&=(\mathcal{L}_{\mathbf{V}}T_{ab\dots ij\dots;c})\omega_{c}+(\mathcal{L}_{\mathbf{V}}T_{ab\dots ij\dots;k})\omega_{k}=0,
\end{align*}
so all covariant derivatives of these tensors are also invariant under the action of $\mathbf{V}$.

Now differentiate the relation $\mathcal{L}_{\mathbf{V}}\omega_{ab}=0$:
\[
\mathcal{L}_{\mathbf{V}}(d\omega_{ab})=\tfrac{1}{2}(\mathcal{L}_{\mathbf{V}}S_{abcd})\omega_{c}\wedge\omega_{d}=0,
\]
so \emph{for an isometry interpreted as a Riemannian submersion, all of the invariants $S_{ijkl}$, $S_{abcd}$, $M_{ija}$, $K_{iab}$ and all of their covariant derivatives are invariant under any of the Killing vector fields.}

\subsection{Condition for a Riemannian submersion to be an isometry}
\label{sec:necess-suff-cond}
We have learned that the condition that all differential invariants and their covariant derivatives are invariant under a vector field is the necessary condition for the vector field, which is along a submersion direction, to be an isometry. This condition is also sufficient: it suffices to note that the Riemann tensor and their derivatives of the whole space can be reconstructed by using all these invariants and their derivatives, with \eqref{eq:oldriemcurv}. Then according to the theory of equivalence, since the differential invariants of the whole space match up to all orders, the group action generated by the vector field is a symmetry of the theory, and symmetry in this theory is exactly isometry.

However, checking equality of differential invariants to all orders is impractical and unnecessary. Since we already have a submersion, we only need to ensure the equality of $\omega_{i}$, $\omega_{a}$, $\pi_{ij}$, $\omega_{ab}$ to its copy under the vector field. Then according to the general theory of equivalence, since in the bundle we do not have any excessive symmetry group at our disposal and the Frobenius theorem is sufficient, we only need to check the vanishing under Lie derivative of the invariants that appear directly in equation \eqref{eq:redrsreq}. Hence, in addition to $M_{ija}$, $K_{iab}$, $S_{abcd}$, $S_{ijkl}$, we need to check the vanishing of the Lie derivative of $K_{ia[b;c]}$ and $M_{ij[a;b]}$. The vanishing of the Lie derivatives of these quantities hence constitute the necessary and sufficient condition for a vector which already generates parts of a Riemannian submersion to be an isometry.

A special case occurs when we have a single vector field, then since we only have a one co-dimensional foliation, $K_{ia[b;c]}$ and $M_{ij[a;b]}$ vanish identically, as well as $S_{abcd}$. As $S_{ijkl}$ is invariant under the vector field automatically (condition for Riemannian submersion), we only need to check the vanishing of Lie derivatives of $M_{ija}$ and $K_{iab}$. But now we have $\mathbf{V}=\lambda \mathbf{I}_{0}$ ($\mathbf{I}_{0}$ is the single tangent vector along the leaves and $\lambda$ is a positive scalar function), and
\[
\mathcal{L}_{\mathbf{V}}M_{ij}=\lambda \dot M_{ij},\qquad \mathcal{L}_{\mathbf{V}}K_{i}=\lambda\dot K_{i},
\] 
where we have suppressed all $0$ indices and used a dot to denote covariant derivation in the fibre direction. Hence, we require $\dot M_{ij}=0$ and $\dot K_{i}=0$, \emph{and in this co-dimension $1$ case, in general when we want to check a quantity is invariant under the submersion vector field, we only need to check that its covariant derivative in this direction vanishes}.

We can also integrate to obtain the parameter $\lambda$ as a function of the coordinates. \emph{Working with a section of the principal bundle}, we know that our vector field $\mathbf{v}=\lambda\mathbf{I}_{0}$ already satisfies
\[
\mathcal{L}_{\lambda\mathbf{I}_{0}}\sum\omega_{i}\otimes\omega_{i}=0.
\]
With our condition $M_{ija}=-M_{jia}$, $K_{iab}=K_{iba}$, this is an identity. Hence we only need to require
\begin{align*}
  0&=\mathcal{L}_{\lambda\mathbf{I}_{0}}(\omega_{0}\otimes\omega_{0})\\
  &=2\dot \lambda\,\omega_{0}\otimes\omega_{0}+2(\lambda_{;i}-\lambda K_{i})(\omega_{i}\otimes_{S}\omega_{0}),
\end{align*}
so the positive function $\lambda$ must also be constant along the leaves. As for $\lambda_{;i}=\lambda K_{i}$, using any coordinates where $x^{i}$ are the coordinates on the reduced manifold, it suffices to integrate the equation
\[
\frac{\pd\log\lambda}{\pd x^{i}}=K_{i}.
\]

If we do not yet know the condition for isometry, we can see it from this equation: $K_{i}$ is the derivative on the reduced manifold of something independent of the fibre coordinates, so $\dot K=0$. Since $K_{i}$ is obtained from a differential, $K_{[i;j]}=0$, but we have from the general equations of submersion, $K_{[i|ab;|j]}=-M_{ij(a;b)}$, so in this case $\dot M_{ij}=-K_{[i;j]}=0$.

On the other hand, if the co-dimension is greater than one, attempting to carry out the same explicit integration yields complicated partial differential equations containing higher derivatives.

\part*{Applications to relativistic fluid mechanics}
\addcontentsline{toc}{part}{Applications to relativistic fluid mechanics}

In this part we will apply the framework of structure-preserving submersions to study relativistic flows. As we have mentioned, the emphasis of this approach is on the reduction of variables.

\section{Born rigid flow}
\label{sec:relat-born-rigid}

\subsection{Definition and structure}
\label{sec:notion-born-rigid}\index{Born rigid flow}

The notion of rigidity in Newtonian spacetime is intuitive and straightforward. It is natural to extend this notion to the theory of relativity discovered by Einstein. Such a definition is given by Born \cite{born1909}, which reads:

\begin{dfn}
  {A body is called rigid if the distance between neighbouring pair of particles, measured orthogonal to the worldlines of either of them, remains constant along the worldline.}
\end{dfn}

Let us immediately note a few things. First, the same wording can be used to define rigidity in Newtonian spacetime, if the notion of worldline and orthogonality are defined in the obvious manner. Second, in relativity, \emph{this condition should be taken to be infinitesimal}, since the distance ``orthogonal'' to a worldline only has a precise meaning in such a limit (in Newtonian spacetime, however, this condition makes sense even with respect to finite distance). If we denote the vector field along the worldlines to be $\lambda\mathbf{I}_{0}$ and an orthonormal co-frame to be $\omega_{0}$, $\omega_{i}$, we see that this condition for rigidity is simply
\[
\mathcal{L}_{\lambda\mathbf{I}_{0}}(\omega_{i}\otimes\omega_{i})=0,
\]
and comparing with our discussion of semi-Killing vectors, we see that this condition is just the condition for a Riemannian submersion of codimension one. Hence we can use our results about Riemannian submersions to study rigid flow in relativity (of course, since we have been doing Riemannian submersions, we need the so-called ``Wick rotation'' trick, which we will apply implicitly).

The structural equations of rigid flow in relativity is a much simplified version of the structural equations for Riemannian submersion. Let $\omega_{0}$, $\omega_{i}$ be the basic coframe, parts of the vertical forms decomposes:
\[
\omega_{0i}=-\omega_{i0}=K_{i00}\omega_{0}-M_{ij0}\omega_{j}\equiv K_{i}\omega_{0}-M_{ij}\omega_{j},
\]
and rigidity requires $M_{(ij)}=0$. Now we can immediately give physical interpretations to $M_{ij}$ and $K_{i}$: since we have
\[
d\mathbf{I}_{0}=K_{i}\omega_{0}\mathbf{I}_{i}+M_{ij}\omega_{j}\mathbf{I}_{i},
\]
we see that $K_{i}$ and $M_{ij}$ are just the acceleration and vorticity of the flow respectively, and the flow is shear-free and expansion-free (for an introduction to the usual definition of these quantities using coordinates, see \cite{ehlers}). The modified connection is defined by
\[
\pi_{ij}=\omega_{ij}-M_{ij}\omega_{0},
\]
and the structural equations now read
\[
\left\{
  \begin{aligned}
    d\omega_{i}&=-\pi_{ij}\wedge\omega_{j},\\
    d\omega_{0}&=-K_{i}\omega_{0}\wedge\omega_{i}-M_{ij}\omega_{i}\wedge\omega_{j},\\
    d\pi_{ij}&=-\pi_{ik}\wedge\pi_{kj}+\tfrac{1}{2}S_{ijkl}\omega_{k}\wedge\omega_{l}.
  \end{aligned}
\right.
\]
We can also write the Riemann tensors in terms of the invariants:
\begin{equation}
  \label{eq:bornstreqns}
  \left\{
  \begin{aligned}
    R_{ijkl}&=S_{ijkl}+M_{il}M_{jk}-M_{ik}M_{jl}-2M_{ij}M_{kl},\\
    R_{ijk0}&=M_{ij;k}-M_{jk}K_{i}+M_{ik}K_{j}+M_{ij}K_{k},\\
    R_{0i0j}&=M_{ik}M_{jk}-K_{(i;j)}-K_{i}K_{j},
  \end{aligned}
\right.
\end{equation}
and for the algebraic relations, in addition to those that involve exchanges of derivation indices, the usual ones for Riemann tensors, and those that are obtained by covariant differentiation, we have
\[
\left\{
  \begin{aligned}
    M_{[ij;k]}&=3M_{[ij}K_{k]},\\
    M_{ij;0}&=-K_{[i;j]}.
  \end{aligned}
\right.
\]
There are no special equations for second order derivations---this is in contrast with the higher codimensional case. We can write the table of the involutive seeds
\begin{center}
  \begin{tabular}{cc}
    Invariant&Independent terms\\
    \hline
    $M_{ij}$&$i>j$\\
    $K_{i}$&all\\
    $S_{ijkl}$&$i>j,\ k>l,\ i\ge k,\ j\ge l$\\
    \hline
    $M_{ij;k}$&$i>j,\ i\ge k$\\
    $K_{i;0}$&all\\
    $K_{i;j}$&all\\
    $S_{ijkl;m}$&$i>j,\ k>l,\ i\ge k,\ j\ge l,\ k\ge m$\\
  \end{tabular}
\end{center}
If the space is $n$ dimensional, $s_{n}=n-1$. Except for $n\le 2$, this is less than the degree of freedom of a Riemannian space, $n(n-1)/2$, showing that not all spaces admit rigid flow. Indeed, these are just special cases of results already obtained for general Riemannian submersions.

\subsection{Self-gravitating perfect fluid under dissipationless flow}
\label{sec:self-grav-perf}\index{perfect fluid}

As a first example of how such a system can be used for real problems,
we  now study the degree of freedom of a dissipationless flow of perfect fluid. A flow is dissipationless if and only if it is shear-free and expansion-free, so by our interpretation of $M_{ij}$ a dissipationless flow is just a Born rigid flow. A fluid is perfect if its energy momentum tensor is of the expression
\[
T_{\mu\nu}=(\rho+p)v_{\mu}v_{\nu}+p\delta_{\mu\nu},
\]
where $p$ is the (isotropic) fluid pressure and $\rho$ is the fluid energy density. The fluid is self-gravitating if this energy momentum tensor is coupled to the Einstein equation. The Einstein equation reads
\[
R_{\mu\nu}-\tfrac{1}{2}R\delta_{\mu\nu}=T_{\mu\nu}
\]
(we could add a cosmological constant but this just amounts to shifting  the pressure and energy density). In our situation, this gives
\[
\left\{
\begin{aligned}
  T_{00}{}={}&{}-\tfrac{1}{2}S-\tfrac{3}{2}M_{ij}M_{ij}=\rho+p,\\
  T_{ij}{}={}&{}(S_{ij}-\tfrac{1}{2}\delta_{ij}S)-(K_{(i;j)}-\delta_{ij}K_{k;k})-(K_{i}K_{j}-\delta_{ij}K_{k}K_{k})\\
  &{}+\tfrac{1}{2}(4M_{ik}M_{kj}+4M_{kl}M_{kl})=\delta_{ij}p,\\
  T_{i0}{}={}&=M_{ji;j}-2M_{ij}K_{j}.
\end{aligned}
\right.
\]
(The energy-momentum conservation is automatic since we couple it to Einstein gravity).
We interpret these equations in the following way: we have added two variables to our system, namely $\rho$ and $p$. The second equation above sets the Ricci tensor of the reduced space $S_{ij}$ to functions of the other variables. The first equation gives the energy density $\rho$ in terms of the other variables. Hence now we have the table
\begin{center}
  \begin{tabular}{cc}
    Invariant&Normal terms\\
    \hline
    $M_{ij}$&$i>j$\\
    $K_{i}$&all\\
    $S_{ijkl}$&after specifying Ricci tensor\\
    $p$&all\\
    \hline
    $M_{ij;k}$&$i>j,\ i\ge k,\ i,k$ not both maximal\\
    $K_{i;0}$&all\\
    $K_{i;j}$&all\\
    $S_{ijkl;m}$&after specifying Ricci tensor\\
    $p_{;i}$&all\\
    $p_{;0}$&all
  \end{tabular}
\end{center}
The degree of freedom is now $s_{n}=n$, coming from $K_{i;0}$ and $p_{;0}$.

This result also suggests how we can specify the Cauchy data for such a problem. For example, we can specify completely the acceleration $K_{i}$ and the pressure $p$ in terms of the rest. Then
\begin{center}
  \begin{tabular}{cc}
    Invariant&Normal terms\\
    \hline
    $M_{ij}$&$i>j$\\
    $S_{ijkl}$&after specifying Ricci tensor\\
    \hline
    $M_{ij;k}$&$i>j,\ i\ge k,\ i,k$ not both maximal\\
    $S_{ijkl;m}$&after specifying Ricci tensor\\
  \end{tabular}
\end{center}
Now $s_{n}=s_{n-1}=0$ and $s_{n-2}=n^{2}-3n-1$, and this is the number of functions we need to specify on the Cauchy hypersurface in order to completely integrate the system.

Note that if we specify in addition an \emph{equation of state} which gives $\rho$ in terms of $p$, then in the table of involutive seeds $p_{;i}$ and $p_{;0}$ disappears, and the degree of freedom is only $s_{n}=n-1$. The Cauchy data is now a specification of $K_{i}$ only, the acceleration of the flow.

\subsection{Rigid flow in homogeneous spacetime}
\label{sec:rigid-flow-homog}\index{Herglotz-Noether theorem}

Our next application begins with the problem of the existence of rigid flow in Minkowski spacetime. Since there is no essential difference, we should at the same time include the study of rigid flow in all homogeneous time, i.e., de Sitter and anti-de Sitter spacetime. Our result would generalise the following classical theorem to all dimensions and to all homogeneous spacetimes (and later to all conformally flat spacetimes of dimension $\ge 4$):

\begin{thm*}
[Herglotz--Noether] \emph{In the spacetime of $3+1$ special relativity, every rotational Born-rigid flow must be isometric.}  
\end{thm*}

See \cite{herglotz1910,noether1910,noether1910,Giulini:2006p66} for more details about the proof of the classical theorem.

Specifying the geometry of the total spacetime amounts to specifying the quantities $R_{ijkl}$, $R_{ijk0}$ and $R_{0i0j}$ in \eqref{eq:bornstreqns}. 
Since the total spacetime is now assumed to be homogeneous, these quantities are constant. In particular, they do not depend on the fibre coordinates. Then we can immediately see that $M_{ij}$ also does not depend on the fibre coordinates: it suffices to take the first equation of \eqref{eq:bornstreqns}
\[
R_{ij\bar i\bar j}=S_{ij\bar i\bar j}-3M_{ij}M_{\bar i\bar j}
\]
where $\bar i$ and $i$ represent the same index, with no summation over them.

Now there are two cases that has to be discussed separately.

\paragraph{First case.} Assume that $M_{ij}$ does not vanish identically. Then the second equation contains the equations
\[
R_{ij\bar i0}=M_{ij;\bar i}+2M_{ij}K_{\bar i},
\]
since $M_{ij}$ does not depend on the fibre coordinates, $M_{ij;k}$ does not neither. Then the above equation shows that if $M_{ij}\neq 0$, then $K_{k}$ does not depend on the fibre coordinates where $k$ can take any index that appears in non-vanishing $M_{ij}$. Now suppose $l$ is an index that does not appear in the index of any non-vanishing $M_{ij}$, then for a certain non-vanishing $M_{ij}$, for example $M_{12}$, we have
\[
R_{12l0}=M_{12;l}+M_{12}K_{l},
\]
showing that $K_{l}$ does not depend on the fibre coordinates either.

Now we see that both $M_{ij}$ and $K_{k}$ does not depend on the fibre coordinates. By our discussion of the relation of semi-Killing vector fields to Killing vector fields,, we see that the submersion is actually generated by a Killing vector field. It is easy to see that in order to ensure $M_{ij}\neq 0$, this Killing vector field must contain some rotational part.

Since specifying a Killing vector field in a homogeneous space it suffices to specify a few constants at a point, in this case
\[
s_{1}=s_{2}=\dots=0.
\]

\paragraph{Second case.} Now assume $M_{ij}=0$ identically. The second equation immediately gives $R_{ijk0}=0$, hence except for the case where the total dimension is $2$, the space must be a Minkowski space for this case to arise. Granted this, then the first equation gives $S_{ijkl}=0$: the reduced space is flat. The third equation gives the equation
\[
K_{(i;j)}=-K_{i}K_{j}.
\]
Now let us use the theory of involutive seeds \cite{hu1} to study this system. All invariants involving $M_{ij}$ or $S_{ijkl}$ now vanish. For those involving $K_{i}$, we have, in addition to the equation above,
\[
K_{[i;j]}=-M_{ij;0}=0,
\]
so the only independent quantities are now
\[
K_{i},\qquad K_{i;0},\qquad K_{i;00},\qquad\dots
\]
hence
\[
s_{1}=n-1,\qquad s_{2}=s_{3}=\dots=0,
\]
the general solution depends on $n-1$ functions of $1$ variable.

The case where the total dimension is two gives instead
\[
K_{1;1}=-K_{1}K_{1}-R_{0101},
\]
and
\[
s_{1}=1,\qquad s_{2}=0.
\]

Note that this reasoning cannot be extended to the higher codimensional case: if we mirror the reasoning, we only get $\sum_{a}M_{ija}M_{\bar i\bar j a}$ independent of the fibre coordinates. Roughly speaking, the proof goes through because on $\rs$ there is no non-trivial connected isotropy group.

It is possible to show that, in the generic case, after specifying completely the geometry of the total space, the degree of freedom of Born-rigid motion  is then zero, and there may not exist any rigid motion at all: see \cite{hu1}.

\paragraph{Geometrical interpretations: first case}
\label{sec:geom-interpr}
Here we simply have a rotational Killing vector on spacetime. Since $M_{ij}\neq 0$, the distribution $\omega_{0}=0$ gives
\[
d\omega_{0}=-M_{ij}\omega_{i}\wedge\omega_{j},
\]
i.e., it is not completely integrable. This shows that it is impossible to find a coordinate system $(\mathbf{x},t)$ on the spacetime such that $t$ is the parameter along each fibre, $\mathbf{x}$ is the parameter on the reduced space and \emph{for every constant $t$ section we have a section isometric to the reduced space} (such a picture would correspond to our intuition in the Galilean case: the constant time sections are just the ``moving rigid body'').

The equation
\[
S_{ij\bar i\bar j}=R_{ij\bar i\bar j}+3M_{ij}M_{\bar i\bar j}
\]
shows that the reduced space is more positively curved than the spacetime, since $M_{ij}^{2}$ is always positive (this is in the Riemannian sense: in the pseudo-Riemannian case the statement still has some content under a Wick rotation). However, the reduced space is not homogenous. For example, for dimension $3+1$ Minkowski spacetime, we can use a coframe such that $M_{12}\neq 0$ but $M_{13}=M_{23}=0$. Then it is easy to see that
\[
S_{ijkl}=0 \qquad \text{except for} \qquad S_{1212}>0,
\]
and $\omega_{3}$ is actually a flat direction.

It should also be noted that the solutions we obtain are local, and actually in general \emph{global solutions cannot exist}. Indeed, suppose that we choose an inertial frame in a Minkowski spacetime such that the Killing vector consists of pure rotation. Then in this frame the velocity of the vector field is proportional to the distance from the centre of rotation, and as we go further and further this velocity will exceed the speed of light.

\paragraph{Geometrical interpretation: second case} Now $M_{ij}=0$, the distribution $\omega_{0}=0$ is completely integrable: the picture of some rigid body moving in spacetime is valid. Furthermore, since
\[
S_{ijkl}=R_{ijkl},
\]
we know what these moving bodies are: they are just hyperplanes. The non-zero Cartan character comes from the time derivative $K_{i;0}$. This shows that to specify completely the motion, we can take an arbitrary point on the moving body, specify its acceleration $K_{i}$ at a certain instant and the change of its acceleration $K_{i;0}$ at all time. This can be visualised as an ordinary plane moving arbitrarily in three dimensional space, but we need to remember that the ``time'' in this picture is intrinsic and depends only on how the planes at different time are stacked together and independent of the parameter time in our model.

\subsection{Rigid flow in conformally flat spacetime.}\index{Herglotz-Noether theorem}
\label{sec:rigid-flow-conf}

Note that in the above calculations we only used a few of the equations \eqref{eq:bornstreqns}: more precisely, equations whose left hand sides are the following:
\[
R_{ij\bar i\bar j},\qquad R_{ijk0},
\]
and the rest of the equations are easily seen to be satisfied identically. Now we shall investigate a problem of specifying a weaker condition on the total spacetime: the problem for which the spacetime is conformally flat. For this, we first need the expressions for the Ricci tensor and scalar in terms of invariants of submersion, which are easily calculated to be
\[
\left\{
  \begin{aligned}
    R_{ij}&=S_{ij}+2M_{ik}M_{kj}-K_{(i;j)}-K_{i}K_{j},\\
    R_{00}&=-K_{i;i}-K_{i}K_{i}+M_{ij}M_{ij},\\
    R_{0i}&=M_{ji;j}-2M_{ij}K_{j},\\
    R&=S-2K_{i;i}-2K_{i}K_{i}-M_{ij}M_{ij}.
  \end{aligned}
\right.
\]
Now we can form the Weyl tensor, by subtracting various traces from the Riemann tensor. The general formula is
\begin{align*}
  W_{\mu\nu\rho\lambda}{}={}&{}R_{\mu\nu\rho\lambda}-\tfrac{1}{n-2}(\delta_{\mu\rho}R_{\lambda\nu}-\delta_{\mu\lambda}R_{\rho\nu}-\delta_{\nu\rho}R_{\lambda\mu}+\delta_{\nu\lambda}R_{\rho\mu})\\
  &{}+\tfrac{1}{(n-1)(n-2)}R(\delta_{\mu\rho}\delta_{\lambda\nu}-\delta_{\mu\lambda}\delta_{\rho\nu}).
\end{align*}
Specialising to our present case, we have
\[
\left\{
  \begin{aligned}
    W_{ijkl}{}={}&{}R_{ijkl}-\tfrac{1}{n-2}(\delta_{ik}R_{lj}-\delta_{il}R_{kj}-\delta_{jk}R_{li}+\delta_{jl}R_{ki})\\
    &{}+\tfrac{1}{(n-1)(n-2)}R(\delta_{ik}\delta_{lj}-\delta_{il}\delta_{kj}),\\
    W_{i0j0}{}={}&{}R_{i0j0}-\tfrac{1}{n-2}(R_{ij}+R_{00}\delta_{ij})+\tfrac{1}{(n-1)(n-2)}R\delta_{ij},\\
    W_{ijk0}{}={}&{}R_{ijk0}-\tfrac{1}{n-2}(\delta_{ik}R_{0j}-\delta_{jk}R_{0i}).
  \end{aligned}
\right.
\]

Now what we do with this mess? First observe that we can divide the Riemann tensor, Ricci tensor and Ricci scalars into two classes. The first class comprises of
\[
R_{ijkl},\qquad R_{0i0j},\qquad R_{ij},\qquad R_{00}, \qquad R
\]
which contain only terms in
\[
M_{ij}M_{kl},\qquad K_{(i;j)}+K_{i}K_{j}\equiv Q_{ij},
\]
and those involving $S_{ijkl}$.
We shall also write $Q= \tr Q_{ij}=K_{i;i}+K_{i}K_{i}$. The second class comprises of
\[
R_{ijk0},\qquad R_{0i},
\]
which contain only terms in
\[
M_{ij;k}, \qquad M_{ij}K_{k}.
\]
Then observe that $W_{ijkl}$ and $W_{i0j0}$ also belongs to the first class, whereas $W_{ijk0}$ belongs to the second class. We can hence mimic our procedure in the homogeneous case, by first using equations of the first class to solve for $M_{ij}$ in terms of the quantities which are independent of the fibre coordinates, and then using equations of the second class to solve for $K_{i}$.

In a sense, the problem is actually easier when the dimension is large: when all four indices of $W_{ijkl}$ are different, it is just the Riemann tensor $R_{ijkl}$. Even though we cannot get nice quadratic terms such as $M_{ij}M_{\bar i\bar j}$ now, the number of independent equations coming from $W_{ijkl}$ is $O(n^{4})$, whereas the number of variables of $M_{ij}$ is only $O(n^{2})$, and they can be solved completely. Then use the equations $W_{ijk0}$ for which all indices are different, which are $O(n^{3})$ in number, to solve for the $O(n)$ terms $K_{i}$. Again this can be solved completely, except for the case where $M_{ij}=0$, as before.

In low dimensions, however, we do not have this luxury, and we need to be more precise of what we do. We will divide our procedure into two steps.

\paragraph{Step 1.}
First let us calculate the quantity $W_{i0j0}$:
\[  W_{i0j0}{}={}{}\tfrac{n}{n-2}(M_{ik}M_{jk}-\tfrac{1}{n-1}\delta_{ij}M_{kl}M_{kl})-\tfrac{1}{n-2}(S_{ij}-\tfrac{1}{n-1}\delta_{ij}S)
  -\tfrac{n-3}{n-2}(Q_{ij}-\tfrac{1}{n-1}\delta_{ij}Q),
\]
which we can write as (since the space is conformally flat, the Weyl tensor vanishes)
\begin{equation}
  \label{eq:eqnforq}
  Q_{ij}-\tfrac{1}{n-1}\delta_{ij}Q=\tfrac{n}{n-3}(M_{ik}M_{jk}-\tfrac{1}{n-1}\delta_{ij}M_{kl}M_{kl})-\tfrac{1}{n-3}(S_{ij}-\tfrac{1}{n-1}\delta_{ij}S).
\end{equation}

On the other hand, $W_{ijkl}$ can be written
\[
W_{ijkl}=R_{ijkl}-\tfrac{1}{n-2}(\delta_{ik}F_{lj}-\delta_{il}F_{kj}-\delta_{jk}F_{li}+\delta_{jl}F_{ki})
\]
where we have defined
\[
F_{ij}=R_{ij}-\tfrac{1}{2(n-1)}\delta_{ij}R.
\]
We can use \eqref{eq:eqnforq} to derive the expression of $F_{ij}$ involving only the curvatures and $M_{ij}$:
\[
F_{ij}=\tfrac{1}{n-3}((n-2)S_{ij}-\tfrac{1}{2}\delta_{ij}S)-\tfrac{3}{n-3}((n-2)M_{ik}M_{jk}-\tfrac{1}{2}M_{kl}M_{kl}\delta_{ij}).
\]
Now investigate the expression of $W_{ijkl}$ with only two distinct indices:
\begin{align*}
  W_{ij\bar i\bar j}{}={}&{}S_{ij\bar i\bar j}-\tfrac{1}{n-3}(S_{i\bar i}+S_{j\bar j})+\tfrac{1}{(n-2)(n-3)}S\\
  &{}-3M_{ij}M_{\bar i\bar j}+\tfrac{3}{n-3}(M_{ik}M_{\bar ik}+M_{jk}M_{\bar j k})-\tfrac{3}{(n-2)(n-3)}M_{kl}M_{kl}.
\end{align*}
Let the indices $i$ and $j$ go through all permutations and sum, we get
\[
\tfrac{n^{2}-4n+5}{(n-2)(n-3)}M_{kl}M_{kl}=(\cdots),
\]
where $(\cdots)$ represents quantities that are independent of the fibre coordinates. Substitute back, we get
\[
3M_{ij}M_{\bar i\bar j}-\tfrac{3}{n-3}(M_{ik}M_{\bar ik}+M_{jk}M_{\bar j k})=(\cdots).
\]
Now let $j$ go through all possible values and sum, we get
\[
\tfrac{6}{n-3}M_{ik}M_{\bar ik}=(\cdots),
\]
and back substitute again gives
\[
-3M_{ij}M_{\bar i\bar j}=(\cdots),
\]
which says that \emph{the quantities $M_{ij}$ are independent of the fibre coordinates}.

\paragraph{Step 2.}
The expression for $W_{ijk0}$ is
\begin{align*}
  W_{ijk0}{}={}&{}M_{ij;k}-\tfrac{1}{n-2}(\delta_{ik}M_{lj;l}-\delta_{jk}M_{li;l})\\
  &{}-M_{jk}K_{i}+M_{ik}K_{j}+M_{ij}K_{k}+\tfrac{2}{n-2}\delta_{ik}M_{jl}K_{l}-\tfrac{2}{n-2}\delta_{jk}M_{il}K_{l},
\end{align*}
and we now know that the first line contain only quantities independent of the fibre coordinates. Consideration of the term $W_{ij\bar i0}$ gives
\[
2M_{ij}K_{\bar j}-\tfrac{2}{n-2}M_{il}K_{l}=(\cdots),
\]
and summing over all $j$ gives
\[
-\tfrac{2}{n-2}M_{il}K_{l}=(\cdots),
\]
back substituting gives
\[
2M_{ij}K_{\bar j}=(\cdots),
\]
so if $M_{ij}\neq 0$ for a certain pair $i$, $j$, then for these two values $K_{i}$ and $K_{j}$ are both independent of the fibre coordinates. If $i$, $j$ and $k$ are all distinct, then $W_{ijk0}$ is just $R_{ijk0}$. Let $k$ be an index such that $M_{lk}=0$ for all choices of $l$, and choose a pair of indices $i$, $j$ such that $M_{ij}\neq 0$. Then consideration of $W_{ijk0}$ gives
\[
M_{ij}K_{k}=(\cdots),
\]
hence \emph{as long as $M_{ij}\neq 0$ for any component, $K_{i}$ is independent of the fibre coordinates}. We have successfully proved the following:

\begin{thm*}
{Any rotational rigid flow in a conformally flat spacetime is a Killing flow.}  
\end{thm*}

Note that unlike the homogenous case, here the result does not assert the existence of any rigid flow, due to the fact that a conformally flat spacetime does not necessarily admit any rotational Killing vector field.

\paragraph{Cases of $n=2, 3$.}
The above proof holds only for the dimension of the spacetime $n\ge 4$: the factors $(n-1)$, $(n-2)$ and $(n-3)$ appear in the numerator at various places. At a deeper level, for dimension less than $4$, the Weyl tensor is trivial. For $n=2$, we know that all spaces are conformally flat, so this case has already been studied in {exist-struct-pres-1}. For $n=3$, the condition for a spacetime to be conformally flat is that the Cotton tensor
\[
C_{\mu\nu\rho}=R_{\mu\nu;\rho}-R_{\mu\rho;\nu}+\tfrac{1}{2(n-1)}(\delta_{\mu\rho}R_{;\nu}-\delta_{\mu\nu}R_{;\rho})
\]
vanishes. This condition involves higher derivatives of $M_{ij}$ and $K_{i}$, and in general does not imply Killing vector fields under rigid flow.

\section{Shear-free rigid flow}
\label{sec:weyl-rigid-flow}\index{Weyl rigid flow}

\subsection{From shear-free flow to Weyl geometry.}
\label{sec:from-conformal-rigid}
In dealing with fluids we often need to consider flows that are shear-free but may have non-zero expansion. The methods of Riemannian submersion are not directly applicable to this case, but can we conceptualise it as some other kind of structure-preserving submersion so that our general method still applies? For any flow, once we separate the flow direction as distinguished, the adapted coframe has decomposition
\[
\omega_{0i}=K_{i}\omega_{0}-M_{ij}\omega_{j}-B_{ij}\omega^{j}-E\omega^{i},
\]
where $K_{i}$, $M_{ij}$, $B_{ij}$, $E$ has respective interpretation acceleration, vorticity, shear and expansion. So the condition for a flow to be shear-free is simply that $B_{ij}=0$. Now working on the base, we can check that a shear-free flow preserves the horizontal part of the metric up to scale \footnote{We can take the vector field to be $\lambda\mathbf{I}_{0}$ for any function $\lambda>0$ instead of simply $\mathbf{I}_{0}$ if we want: it only contributes an overall factor in all the following and does not change any of our conclusions. If the terms to be differentiated by the Lie derivation contains the form $\omega_{0}$, the factor $\lambda$ will have significance, as we have seen before.}:
\[
\mathcal{L}_{\mathbf{I}_{0}}\left(\sum\omega_{i}\otimes\omega_{i}\right)=E(\omega_{i}\otimes\omega_{i}).
\]

For easier calculation, let us lift this unto the bundle and try to find out the lifting condition for a vector field
\[
\mathbf{V}=\mathbf{I}_{0}+V_{ij}\mathbf{I}_{ij}
\]
to satisfy
\[
\mathcal{L}_{\mathbf{V}}\omega_{i}\propto \omega_{i}.
\]
We have
\begin{align*}
  \mathcal{L}_{\mathbf{V}}\omega_{i}&=\mathbf{V}\iprod(-\omega_{ij}\wedge\omega_{j}-M_{ij}\omega_{j}\wedge\omega_{0}-E\omega_{i}\wedge\omega_{0})\\
  &=E\omega_{i}-(V_{ij}+M_{ij})\wedge\omega_{j},
\end{align*}
hence the uplifting is
\[
\mathbf{V}=\mathbf{I}_{0}-M_{ij}\mathbf{I}_{ij}.
\]

The condition in the bundle
\[
\mathcal{L}_{\mathbf{V}}\omega_{i}=E\omega_{i}
\]
is still not very convenient to work with. Let us try to find some quantity that vanishes under the Lie derivative in the bundle. It is reasonable to try the scaling $\theta_{i}=e^{-\Lambda}\omega_{i}$. We have
\begin{align*}
  \mathcal{L}_{\mathbf{V}}\theta_{i}&=\mathcal{L}_{\mathbf{V}}(e^{-\Lambda})\omega_{i}+e^{-\Lambda}E\omega_{i}\\
  &=(E-\mathbf{I}_{0}(\Lambda))\theta_{i},
\end{align*}
If we write $\mathbf{I}_{0}=\pd/\pd t$, it  suffices to integrate the equation
\[
\frac{d\Lambda}{dt}=E
\]
along the flow.
This is the simplest differential equation that one could possibly have, and is always solvable on each flowline, the solution depending on one constant. On the space itself, a solution for such an equation hence depend on a function of $n-1$ parameters, in other words, on any hypersurface transverse to the flowlines we can choose these constants of integration arbitrarily (subject to the appropriate smoothness conditions, of course).

To obtain the complete set of conditions, we need, as usual, to differentiate our condition. We have
\begin{align*}
  0=\mathcal{L}_{\mathbf{V}}d\theta_{i}&=\mathcal{L}_{\mathbf{V}}(-\omega_{ij}\wedge\theta_{j}-e^{\Lambda}M_{ik}\theta_{k}\wedge\theta_{0}-e^{\Lambda}E\theta_{i}\wedge\theta_{0}-d\Lambda\wedge\theta_{i})\\
  &=-\mathcal{L}_{\mathbf{V}}(-\omega_{ij}+e^{\Lambda}M_{ik}\theta_{0})\wedge\theta_{j}-\mathcal{L}_{\mathbf{V}}(d\Lambda-e^{\Lambda}E\theta_{0})\wedge\theta_{i}.
\end{align*}
Hence the forms \footnote{We really need to assume the Lie derivatives are linear in $\theta_{i}$ and apply some manipulation to get the following. We omit the manipulations, which should be familiar now.}
\[
\pi_{ij}\equiv \omega_{ij}-e^{\Lambda}M_{ik}\theta_{0},\qquad \varpi\equiv d\Lambda-e^{\Lambda}E\theta_{0}
\]
are independent of the fibre coordinates. It is also easy to see that
\[
\theta_{i},\qquad \pi_{ij},\qquad \varpi
\]
provide a Weyl connection on the reduced space. \emph{Hence a shear-free flow can be interpreted as a structure-preserving submersion: a submersion that preserves the Weyl structure of the subspace.}

Note that the most general conformal geometry is not what we want: here it contains, in addition to rotations and scaling, other local symmetries: the special conformal transformations. If we include special conformal transformations, the geometry is no longer reductive, and hence there is no way we can define any covariant derivatives along directions on the base manifold---every derivative necessarily leads us upstairs, into the bundle.

\subsection{Riemannian geometry disguised as Weyl geometry.}
\label{sec:riem-geom-disg}

As we have seen,
it is impossible to talk about the reduction of a Riemannian geometry to a Weyl geometry since Weyl geometry is more general than Riemannian geometry. We can remedy this by putting the Riemannian geometry into the form of a Weyl geometry, and talk instead of the reduction of a Weyl geometry to another Weyl geometry. We take a coframe in the Riemannian geometry $\theta^{\mu}$, $\theta^{\mu}{}_{\nu}$ and its curvature $\mathcal{R}^{\mu}{}_{\nu\rho\lambda}$, and do the scaling
\[
\omega^{\mu}=e^{-\Lambda}\theta^{\mu}
\]
for a function $\Lambda$ defined on the base.
The structural equation then becomes
\[
\left\{
  \begin{aligned}
    d\omega^{\mu}&=-\theta^{\mu}{}_{\nu}-d\Lambda\wedge\omega^{\mu},\\
    d\theta^{\mu}{}_{\nu}&=-\theta^{\mu}{}_{\lambda}\wedge\theta^{\lambda}{}_{\nu}+\tfrac{1}{2}e^{2\Lambda}\mathcal{R}^{\mu}{}_{\nu\rho\lambda}\omega^{\rho}\wedge\omega^{\lambda},
  \end{aligned}
\right.
\]
we see that
\[
\omega^{\mu}{}_{\nu}=\theta^{\mu}{}_{\nu},\qquad \tau=d\Lambda,\qquad R^{\mu}{}_{\nu\rho\lambda}=e^{2\Lambda}\mathcal{R}^{\mu}{}_{\nu\rho\lambda}.
\]
In particular, this implies
\[
d\tau=d^{2}\Lambda=0,\qquad F_{\mu\nu}=0.
\]
This condition is also sufficient for the local reducibility of a Weyl geometry to a Riemannian geometry: it suffices to integrate these equations back to get the value of $\Lambda$. Note that for such a geometry, the invariants has \emph{exactly the same} structure as a Riemannian geometry (they have the same value up to scaling), and hence the degree of freedom is exactly the same. In Weyl geometry we have removed one degree of freedom from the metric by scaling, but for the existence of the covariant derivative, we have added a scale connection. In the reducible case, this added scale connection, as we have just seen, is the differential of a function, hence we need to add back one degree of freedom. In other words, a Weyl geometry reducible to Riemannian geometry amounts to taking away one degree of freedom from the metric and add it to somewhere else, and the net change for the degree of freedom is zero (but such a manipulation is not completely in vain: the symmetry group is now one dimension larger).  This should come as a relief for us: we want to study Riemannian geometry, and if the degree of freedom is not the same, we have introduced or removed degree of freedom.

Then there is the question of what happens for $n=2$, surfaces. It is a celebrated result that all two dimensional surfaces are conformally equivalent, so the degree of freedom is zero, definitely not the same as a two dimensional Riemannian surface. The answer is that Weyl geometry for $n=2$ is not the same as the problem of conformal equivalence of surfaces. To see this, let us apply the equivalence method from scratch. Let $\theta^{1}$, $\theta^{2}$ be an orthogonal frame on the surface. For two surfaces to be conformally equivalent, we require
\[
\begin{pmatrix}
  \bar\theta^{1}\\
  \bar\theta^{2}
\end{pmatrix}
=L
\begin{pmatrix}
  \cos t&\sin t\\
  -\sin t&\cos t
\end{pmatrix}
\begin{pmatrix}
  \theta^{1}\\
  \theta^{2}
\end{pmatrix}.
\]
Let the lifted frame be 
\[
\begin{pmatrix}
  \omega^{1}\\
  \omega^{2}
\end{pmatrix}
=L
\begin{pmatrix}
  \cos t&\sin t\\
  -\sin t&\cos t
\end{pmatrix}
\begin{pmatrix}
  \theta^{1}\\
  \theta^{2}
\end{pmatrix},
\]
we can derive the structural equation
\[
\left\{
  \begin{aligned}
    d\omega^{1}&=d(\log L)\wedge\omega^{1}+dt\wedge\omega^{2}+a\omega^{1}\wedge\omega^{2},\\
    d\omega^{2}&=-dt\wedge\omega^{1}+d(\log L)\wedge\omega^{2}+b\omega^{1}\wedge\omega^{2},
  \end{aligned}
\right.
\]
where $a$ and $b$ are torsion. At the linear level, the solution
\[
d(\log L)=l_{1}\omega^{1}+l_{2}\omega^{2},\qquad dt=m^{1}\omega^{1}+m^{2}\omega^{2}
\]
contains four variables $l_{1}$, $l_{2}$, $m_{1}$, $m_{2}$, two of which must be used to set the torsion to zero. Hence the number of free functions is $2$. On the other hand, the Cartan characters have
\[
s_{1}=2,\qquad s_{2}=0,
\]
and since $1\cdot 2+2\cdot 0=2$, \emph{this system, which is not completely integrable, is involutive}, there are no hidden conditions for the existence of integral varieties! The solution depends on $2$ functions of $1$ variables, and the system has an infinite dimensional symmetry group. In this case we are not justified to prolong the system. If we prolong as we did for the higher dimensional case, we get Weyl geometry for $n=2$, but it describes a different geometry.

What we need to take away from this consideration is that in the so-called Weyl form, it is easy to deduce that the invariants $R_{\mu\nu\rho\lambda}$ and, if we do a reduction under a flow, the invariants $M_{ij}$, $K_{i}$ and $E$, are just the scaled counterparts of the quantities in the Weyl frame. \emph{In particular}, going from Riemannian geometry to the Weyl framework does not alter the vertical forms $\omega_{\mu\nu}$ in any way.

\subsection{The structure of shear-free flow.}
\label{sec:invar-riem-geom}

We can draw the following commutative diagram for what we have done so far:
\[
\xymatrix{
M\times SO(n)\ar[r]\ar[d]&M\ar[d]&\text{Riemannian geometry}\ar@/^/[d]^{\text{prolongation}}\\
\ar[r]\ar[u]M\times SO(n)\times \rs^{+}&M\ar[d]\ar[u]&\text{Weyl geometry ($F_{\mu\nu}=0$)}\ar@/^/[u]^{\text{reduction}}\ar@/_/@{.>}[d]_{\text{reduction}}\\
\ar[u]M\times SO(n-1)\times \rs^{+}\ar[r]\ar[d]&M\ar[d]\ar[u]&\text{Weyl geometry (sub-bundle)}\ar@/_/[u]_{\text{inclusion}}\ar[d]^{\text{submersion}}\\
B\times SO(n-1)\times\rs^{+}\ar[r]&B&\text{Weyl geometry}\\
}
\]
the dotted line just means that we have no use of the map indicated. This diagram is curious: on the top row, we have $M\times SO(n)$, but at the bottom row we have $B\times SO(n-1)\times \rs^{+}$: by some magic, we have conjured up the $\rs^{+}$ degree of freedom from air. The equation \eqref{eq:magickiller} tells us that, in this case, since $F_{\mu\nu}=0$,
\begin{equation}
  \label{eq:eb8}
  \left\{
  \begin{aligned}
    G_{ij}&=2E_{0}M_{ij}{}^{0},\\
    E_{0;i}&=-E_{0}K_{i}{}^{0}{}_{0}.
  \end{aligned}
\right.
\end{equation}
and if $G_{ij}\neq 0$, we really have magic here: we have derived inhomogeneity in the $\rs^{+}$ part from a geometry where no $\rs^{+}$ part exists, recalling that the group $SO(n-1)\times\rs^{+}$ is \emph{not} a subgroup of $SO(n)$. Observe that to realise this, the restrictions are huge: not only do we have $G_{ij;0}=0$, so the degree of freedom of $G_{ij;0}$ lies only in the reduced space, but since $G_{[ij;k]}=0$, locally the degree of freedom of $G_{ij}$ is only that of a vector on the reduced space.

Let us note several things. First is that, from the Bianchi identity $G_{ij;0}=0$, we get
\[
E_{0;0}M_{ij}{}^{0}+E_{0}M_{ij}{}^{0}{}_{0;0}=0,
\]
so if $E_{0}\neq 0$, we know how $M_{ij}{}^{0}$ scales on the fibre. From the same equation, we see that $M_{ij}{}^{0}$ is preserved on the fibre up to scale, so we can now write
\[
M_{ij}{}^{0}{}_{0;0}=-\lambda_{0} M_{ij}{}^{0}{}_{0},
\]
then
\[
E_{0;0}M_{ij}{}^{0}-E_{0}\lambda_{0}M_{ij}{}^{0}{}_{0}=M_{ij}{}^{0}(E_{0;0}-E_{0}\lambda_{0})=0,
\]
so if in addition $M_{ij}{}^{0}\neq 0$, the way $E_{0}$ scales on the fibre is related to the way $M_{ij}{}^{0}$ scales on the fibre, given by
\[
E_{0;0}=E_{0}\lambda_{0}.
\]

We also have another formula for $G_{ij}$: \eqref{eq:mb9}, which can now be written
\begin{equation}
  \label{eq:mb10}  
K_{[i|}{}^{0}{}_{0;|j]}=(\lambda_{0}-E_{0})M_{ij}{}^{0}.
\end{equation}

Another nice property is that, since $G_{ij;0}=0$ and $M_{ij0}$ is proportional to $G_{ij}$, if neither $M_{ij0}$ nor $E_{0}$ vanishes the principal bundle can be reduced further: basically $M_{ij0}$ is invariant up to scale on each fibre, so it is a well-defined quantity on the reduced space up to scale. We can then use the residual $SO(n-1)$ symmetry on the reduced space to set various components of $M_{ij0}$ to zero. For example, if the reduced space is only three dimensional, then we can always effect a reduction of the principal bundle such that $M_{130}=M_{230}=0$.

Again we can calculate the following table for the generic case
\begin{center}
  \begin{tabular}{cc}
    Invariant&Normal terms\\
    \hline
    $E_{0}$&all\\
    $M_{ij0}$&$i>j$\\
    $K_{i00}$&all\\
    $S_{ijkl}$&$i>j,\ k>l,\ i\ge k,\ j\ge l$\\
    \hline
    $M_{ij0;k}$&$i>j,\ i\ge k$\\
    $K_{i00;0}$&all\\
    $K_{i00;j}$&all\\
    $S_{ijkl;m}$&$i>j,\ k>l,\ i\ge k,\ j\ge l,\ k\ge m$\\
  \end{tabular}
\end{center}
Except for $E_{0}$ which does not contribute to the Cartan characters, this table is exactly the same as in the Born rigid flow case!

\subsection{Self-gravitating perfect fluid under shear-free flow}
\label{sec:self-grav-perf-1}\index{perfect fluid}

The condition that the spacetime is formed by a self-gravitating perfect fluid amounts to constraints on the Einstein tensors of the total space. In our framework, these constraints amounts to
\[
\left\{
  \begin{aligned}
    \rho&=R^{0}{}_{0}-\tfrac{1}{2}R,\\
    P&=R^{i}{}_{\bar i}-\tfrac{1}{2}R,\\
    0&=R_{ij},& (i\neq j),\\
    0&=R_{0i},
  \end{aligned}
\right.
\]
where $\rho$ and $P$ are the scaled versions of the energy density and pressure, respectively.
Unless we introduce an equation of state by hand, the first equation is useless: it can be thought of the equation of state itself. The second introduces $p-1=n-2$ constraints, the third $p(p-1)/2$ and the fourth $p$ (it can be checked that $R_{ij}=R_{ji}$ and $R_{0i}=R_{i0}$ are identities under our assumption $F_{\mu\nu}=0$).

Using the invariants we can rewrite these equations
\begin{equation}
  \label{eq:merdeface}
\left\{
\begin{aligned}
  \rho{}={}&-\tfrac{1}{2}S+2M^{i}{}_{j0}M_{i}{}^{j0}-\tfrac{1}{2}K^{i0}{}_{0;i}-\tfrac{1}{2}K^{i0}{}_{0}K_{i0}{}^{0}-\tfrac{p}{2}E^{0}\lambda_{0}+\tfrac{p^{2}-p}{2}E^{0}E_{0},\\
  P{}={}&S^{\bar i}{}_{i}-\tfrac{1}{2}S-2M^{\bar ij0}M_{ij0}+M_{jk0}M^{jk0}-K^{\bar i00}K_{i00}\\
  &+\tfrac{p-2}{2}E_{0}\lambda^{0}+\tfrac{(p-1)(p-2)}{2}E^{0}E_{0}-K^{\bar i0}{}_{0;i}+\tfrac{1}{2}K^{j0}{}_{0;j}+\tfrac{1}{2}K^{j00}K_{j00},\\
  0{}={}&\tfrac{1}{2}(S^{i}{}_{j}+S_{j}{}^{i})-M^{ki0}M_{kj0}-K^{i00}K_{j00}-\tfrac{1}{2}(K^{i0}{}_{0;j}+K_{j0}{}^{0;i}),\\
  0{}={}&M^{i}{}_{j0;i}+2M_{ij}{}^{0}K^{i}{}_{00}+pE_{0}K_{j}{}^{0}{}_{0}.
\end{aligned}
\right.
  \end{equation}

This is the subject of the celebrated Ellis conjecture (for a review and partial results, see \cite{ellis1}): namely these constraints together with any barotropic equation of state which satisfies $\rho+P\neq 0$ require either $E_{0}=0$ or $M_{ij}{}^{0}=0$ to hold. Of course, since $G_{ij}=M_{ij0}E^{0}$, this conclusion is equivalent to the statement that the reduced space also has a Riemannian connection.

We can see the difficulty in proving this theorem: even in the lowest dimensions ($n=4$, the usual general relativity framework) there are a lot of variables with complicated relations among them. However, there are many partial results in which additional assumptions are added, of which we will mention just one: if in four dimensions the acceleration of the fluid is zero $(K_{i00}=0)$, then the theorem holds. We shall see that, the proof of this partial result achieves a remarkable simplicity in our approach to the problem, and at the same time we get its extension to higher dimensions. In particular, the proof does not involve writing down any differential equation and is due mainly to the symmetry restrictions of the problems.

Before carrying out with our proof, let us note a trivial result that can be read off immediately from our approach, namely

\begin{prop*}
{For an irrotational fluid $(M_{ij0}=0)$ with vanishing energy flux \footnote{Not necessarily a perfect fluid.}, either the expansion or the acceleration has to vanish.}  
\end{prop*}

Indeed, the vanishing energy flux assumption is
\[
0=R_{0j}=M^{i}{}_{j0;i}+2M_{ij}{}^{0}K^{i}{}_{00}+pE_{0}K_{j}{}^{0}{}_{0},
\]
and the irrotational assumption reduces this to
\[
0=pE_{0}K_{j}{}^{0}{}_{0},
\]
so either $E_{0}=0$ or $K_{j}{}^{0}{}_{0}=0$.\qed

If $K_{i00}=0$ then the fluid is subject to no acceleration, i.e., the flow is geodesic. Now \eqref{eq:mb10} reads
\[
(\lambda_{0}-E_{0})M_{ij}{}^{0}=0.
\]
From now on we shall assume that neither $M_{ij0}$ nor $E_{0}$ vanishes and try to derive a contradiction. The above equation immediately gives
\[
\lambda_{0}=E_{0},
\]
and we have
\[
E_{0;0}=E_{0}E_{0},\qquad M_{ij0;0}=-E_{0}M_{ij0}.
\]

Now take the second equation of \eqref{eq:merdeface}:
\[
  P=S^{\bar i}{}_{i}-\tfrac{1}{2}S-2M^{\bar ij0}M_{ij0}+M_{jk0}M^{jk0}
  +\tfrac{p(p-2)}{2}E^{0}E_{0},
\]
and derive it in the flow direction, recalling that $S_{ijkl;0}=0$:
\[
P_{;0}=4E_{0}M^{\bar ij0}M_{ij0}-2E_{0}M_{jk0}M^{jk0}+p(p-2)E_{0}E_{0}E^{0}.
\]
So far our consideration has been in the vertical directions. Now change our point of view and focus on the horizontal directions, we see that
\[
\sum_{j}M^{\bar ij0}M_{ij0}=\text{quantity transforming trivially under $SO(n-1)$},
\]
and the quantity is the same for all value of the index $i$. A simple manipulation shows that the absolute value of each component
\[
|M_{ij0}|
\]
has the same value, and this quantity is invariant under $SO(n-1)$. As $M_{ij0}$ is antisymmetric in $i$ and $j$, it has to vanish for $n\ge 4$, the cases we consider, contradicting our assumption.\qed

Observe that our proof is valid for all dimensions $n\ge 4$, and we did not require any equation of state: in particular, we did not require $P+\rho\neq 0$. 

It seems likely that for the other cases the equation of state and the dimension restriction are necessary. Note that the equation of state $P=P(\rho)$ gives $P_{;0}=\rho_{;0}P'$, and both sides can be calculated independently by regarding $P'=dP/d\rho$ as a new scalar variable, and these equations do not involve $S_{ijkl}$, so the equation of state is actually quite a lot of new constraints. We shall not pursue the calculations, since we do not yet have new things to add to the existing results in this case.

\subsection{Another look at the conformal Herglotz--Noether theorem}
\label{sec:anoth-proof-theor}\index{Herglotz-Noether theorem}

In all of our above calculations we have taken care not to alter the connection forms $\omega^{\mu}{}_{\nu}$ on the total space, so that the invariants such as $R^{\mu}{}_{\nu\rho\lambda}$ are not changed except for a scaling. But of course we can do otherwise. First take the flat Weyl geometry, with structural equation
\[
\left\{
  \begin{aligned}
  d\omega^{\mu}&=-\omega^{\mu}{}_{\nu}\wedge\omega^{\nu}-\tau\wedge\omega^{\mu},\\
  d\omega^{\mu}{}_{\nu}&=-\omega^{\mu}{}_{\lambda}\wedge\omega^{\lambda}{}_{\nu},\\
  d\tau&=0,    
  \end{aligned}
\right.
\]
and now make a reduction of the bundle into a Riemannian space. Since $d\tau=0$, we can write, locally,
\[
\tau=d\Lambda=\Lambda_{;\mu}\omega^{\mu}.
\] 
for some function $\Lambda$ \emph{defined on the base}. Then
\[
d\omega^{\mu}=-\omega^{\mu}{}_{\nu}\wedge\omega^{\nu}-\Lambda_{;\nu}\delta^{\mu}{}_{\lambda}\omega^{\nu}\wedge\omega^{\lambda},
\]
now the term $-\Lambda_{;[\nu|}\delta^{\mu}{}_{|\lambda]}$ is \emph{torsion} in the Riemannian sense: we have a Riemannian geometry with vanishing curvature but non-vanishing torsion. Refer back to {equivalence-problem-1}, we see that all torsion can be absorbed in Riemannian geometry. After absorption, the space we obtain are \emph{conformally flat}, since it is a scaled version of a flat manifold. 

 It is essential to note that \emph{such an absorption changes the geometry in a fundamental way}: in particular, this is not what has been done in {riem-geom-disg}. There we are concerned with \emph{prolongation}, whereas here the subject is \emph{absorption}.

Now suppose that we have a rigid flow on a conformally flat spacetime. By scaling and absorption, this is equivalent to a Weyl flow on the \emph{totally flat version of the Weyl spacetime}. Then we can use the equations relating the invariants of the total geometry with that of the subgeometry to prove the theorem of {rigid-flow-conf}, the complication for calculation now is that there is the torsion absorption procedure involved (and we need to derive what it means for a vector field to be a Killing vector field in the Riemannian case, interpreted in the Weyl framework). The main difficulty of the proof in {rigid-flow-conf}, on the other hand, is that the Weyl tensor is complicated to calculate. We shall not give more details of the proof using Weyl flow, as it does not really represent a simplification. Using the above reasoning, the theorem proved in {rigid-flow-conf} can be phrased in another way:

\begin{thm*}
{In flat spacetime of dimension $\ge 4$, a rotational conformal rigid flow must be a conformal Killing flow.}  
\end{thm*}

\section{Generic relativistic flow and Newtonian rigid motion}
\label{eq:inexgrf}

From our study of Born rigid flow and Weyl rigid flow it would seem that we have found a pattern, and one is tempted to generalise it further:  formulating arbitrary relativistic flow as a submersion preserving an affine connection, together with a condition on the total affine space requiring that it is derivable from a Riemannian space.

However, this is hardly worth the effort, due to the following reason: note that in our previous examples, the condition \eqref{eq:freeforms} implies \eqref{eq:freeforms2}. On the total space, this implies that the foliation due to the submersion itself is enough to split the tangent space into two parts: only the subspace of each leaf is completely determined. Thus, at least at the linear level, \emph{the foliation on the total space itself is insufficient to determine completely the structure-preserving submersion}: we need some additional data. Of course, in the Riemannian or Weylian case, such problem does not arise, as the metric allows us to canonically determine a complement space.

As the total space is actually derived from a Riemannian space, such a splitting is available to us, but this actually only complicates the matter, as we cannot simply require that the splitting which makes an affine connection available on reduced space is just the splitting due to the Riemannian metric: there is no reason such a splitting will give rise to a structure-preserving submersion, and there is no reason that if this splitting does not give rise to such a submersion, no other splitting will. Furthermore, such a requirement would be physically unjustified. Thus we need to introduce quite a lot of auxiliary variables to parametrise the relationship between the two splittings. Since the whole point of the framework of structure-preserving submersions is reduction of variables, this really defeats the purpose (and the auxiliary functions introduced will in general not have any nice properties).

A case that should be compared and contrasted with this case is Newtonian rigid motion as a structure-preserving submersion. 
Here we formulate the motion on a space with a Galilean connection defined on it. The connection matrix can be written
\[
\begin{pmatrix}
  0&0&0\\
  \tau&0&0\\
  \theta_{i}&\omega_{i}&\omega_{ij}
\end{pmatrix}
\]
with the additional constraints
\[
d\tau=0
\]
which guarantees the existence of absolute time,
\[
d\omega_{ij}=-\omega_{ik}\wedge\omega_{kj}
\]
which guarantees the existence of flat space, and
\[
(d\omega_{i}+\omega_{ij}\wedge\omega_{j})\wedge\theta_{i}=0
\]
which guarantees the absence of velocity-dependent gravitational effect. For how these equations are derived, see \cite{cartan-newtonian}, or the English translation \cite{cartan-n-t}. 
 The structural equation is for the total space is 
\[
\left\{
  \begin{aligned}
    d\tau&=0,\\
    d\theta_{i}&=-\omega_{i}\wedge\tau-\omega_{ij}\wedge\theta_{j},\\
    d\omega_{i}&=-\omega_{ij}\wedge\omega_{j}+\Gamma_{ij}\theta_{j}\wedge\tau+\tfrac{1}{2}\Gamma_{ijk}\theta_{j}\wedge\theta_{k},\\
    d\omega_{ij}&=-\omega_{ik}\wedge\omega_{kj},
  \end{aligned}
\right.
\]
and for the reduced space is
\[
\left\{
  \begin{aligned}
    d\pi_{i}&=-\pi_{ij}\wedge\pi_{j},\\
    d\pi_{ij}&=-\pi_{ik}\wedge\pi_{kj},
  \end{aligned}
\right.
\]
and the differential system is 
\[
\theta_{i}=\pi_{i}.
\]
For the analysis of this system together of its degree of motion, see [...]. Here the forms $\theta_{i}$ is also only defined up to linear combinations of $\tau$. The most crucial difference is that here the group $SO(n)$ is a subgroup of the Galilean group, therefore no funny business of extending the Galilean group to some super group is necessary, and hence we can do calculations using any splitting which completely determines $\theta_{i}$ (which physically amounts to choose an orthonormal frame of reference, not necessarily inertial), and it is not necessary to introduce any auxiliary functions.

\newpage

\addcontentsline{toc}{section}{References}
\bibliographystyle{hplain}

\end{document}